\documentclass[12pt]{article}
\usepackage{amsmath}
\usepackage{amssymb}
\usepackage{color}
\usepackage[unicode,bookmarks,bookmarksopen,bookmarksopenlevel=2,colorlinks,linkcolor=blue,citecolor=green]{hyperref}

\def\comment#1{}
\input{tcilatex}
\begin{document}

\title{\textbf{Multi-Dimensional Conservation Laws\\
and Integrable Systems}}
\author{Z.\,V.~Makridin$^{1,2}$, M.\,V.~Pavlov$^{1,3,4}$ \\
$^{1}${\normalsize {Department of Mechanics and Mathematics,}}\\
[-3.8mm] {\normalsize {Novosibirsk State University,}}\\
[-3.8mm] {\normalsize {Pirogova street 2, 630090, Novosibirsk, Russia}}\\
[-3.8mm] $^{2}${\normalsize {Lavrentiev Institute of Hydrodynamics,}}\\
[-3.8mm] {\normalsize {Lavrentiev street 15, 630090, Novosibirsk, Russia}}\\
[-3.8mm] $^{3}${\normalsize {Sector of Mathematical Physics,}}\\
[-3.8mm] {\normalsize {Lebedev Physical Institute of Russian Academy of
Sciences,}}\\
[-3.8mm] {\normalsize {Leninskij Prospekt 53, 119991 Moscow, Russia}}\\
[-3.8mm] $^{4}${\normalsize {Department of Applied Mathematics,}}\\
[-3.8mm] {\normalsize {National Research Nuclear University MEPHI,}}\\
[-3.8mm] {\normalsize {Kashirskoe Shosse 31, 115409 Moscow, Russia}}}
\maketitle

\begin{abstract}
In this paper we introduce a new property of two-dimensional integrable
systems -- existence of infinitely many local three-dimensional conservation
laws for pairs of integrable two-dimensional commuting flows. Infinitely
many three-dimensional local conservation laws for the Korteweg de Vries
pair of commuting flows and for the Benney commuting hydrodynamic chains are
constructed. As a by-product we established a new method for computation of
local conservation laws for three-dimensional integrable systems. The
Mikhalev equation and the dispersionless limit of the
Kadomtsev--Petviashvili equation are investigated. All known local and
infinitely many new quasi-local three-dimensional conservation laws are
presented. Also four-dimensional conservation laws are considered for
couples of three-dimensional integrable quasilinear systems and for triples
of corresponding hydrodynamic chains.
\end{abstract}

\tableofcontents

\newpage

\begin{flushright}
\textit{dedicated to 70th birthday of Franco Magri}
\end{flushright}

\bigskip

\bigskip

\textit{keywords}: integrable system, Benney hydrodynamic chain,
Kadomtsev--Petviashvili equation, Korteweg de Vries equation,
three-dimensional conservation laws. \bigskip



\newpage

\section{Introduction}

\label{sec:intro}

One of most important properties of \textit{two-dimensional} integrable
systems%
\begin{equation*}
u_{t}^{k}=F^{k}(\mathbf{u},\mathbf{u}_{x},\mathbf{u}_{xx},\mathbf{u}%
_{xxx},\ldots),\text{ \ }k=1,\ldots,N
\end{equation*}%
is an existence of infinitely many local \textit{two-dimensional}
conservation laws%
\begin{equation*}
\bigl(h_{p}(\mathbf{u},\mathbf{u}_{x},\mathbf{u}_{xx},\mathbf{u}%
_{xxx},\ldots)\bigr)_{t}=\bigl(g_{p}(\mathbf{u},\mathbf{u}_{x},\mathbf{u}%
_{xx},\mathbf{u}_{xxx},\ldots)\bigr)_{x}.
\end{equation*}%
In this paper we observe that pairs of \textit{\textbf{two}-dimensional}
integrable commuting flows%
\begin{equation*}
u_{t}^{k}=F^{k}(\mathbf{u},\mathbf{u}_{x},\mathbf{u}_{xx},\mathbf{u}%
_{xxx},\ldots),\text{ \ }u_{y}^{k}=G^{k}(\mathbf{u},\mathbf{u}_{x},\mathbf{u}%
_{xx},\mathbf{u}_{xxx},\ldots)
\end{equation*}%
possess also infinitely many local \textit{\textbf{three}-dimensional}
conservation laws%
\begin{equation*}
\bigl(A_{q}(\mathbf{u},\mathbf{u}_{x},\mathbf{u}_{xx},\ldots)\bigr)_{y}+%
\bigl(B_{q}(\mathbf{u},\mathbf{u}_{x},\mathbf{u}_{xx},\ldots)\bigr)_{t}+%
\bigl(C_{q}(\mathbf{u},\mathbf{u}_{x},\mathbf{u}_{xx},\ldots)\bigr)_{x}=0.
\end{equation*}%
Moreover, obviously, triples of \textit{\textbf{two}-dimensional} integrable
commuting flows%
\begin{equation*}
u_{t}^{k}=F^{k}(\mathbf{u},\mathbf{u}_{x},\mathbf{u}_{xx},\ldots),\text{ \ }%
u_{y}^{k}=G^{k}(\mathbf{u},\mathbf{u}_{x},\mathbf{u}_{xx},\ldots),\text{ \ }%
u_{z}^{k}=Q^{k}(\mathbf{u},\mathbf{u}_{x},\mathbf{u}_{xx},\ldots)
\end{equation*}%
possess infinitely many local \textit{\textbf{four}-dimensional}
conservation laws%
\begin{equation*}
\bigl(A_{q}(\mathbf{u},\mathbf{u}_{x},\ldots)\bigr)_{z}+\bigl(B_{q}(\mathbf{u%
},\mathbf{u}_{x},\ldots)\bigr)_{y}+\bigl(C_{q}(\mathbf{u},\mathbf{u}%
_{x},\ldots)\bigr)_{t}+\bigl(D_{q}(\mathbf{u},\mathbf{u}_{x},\ldots)\bigr)%
_{x}=0.
\end{equation*}%
So, we come to the \textbf{Statement} that any $M-1$ commuting integrable
systems%
\begin{equation*}
u_{t^{m}}^{k}=F_{m}^{k}(\mathbf{u},\mathbf{u}_{x},\mathbf{u}_{xx},\ldots),%
\text{ \ }k=1,2,\ldots,N,\text{ \ }m=2,\ldots,M
\end{equation*}%
possess infinitely many local \textit{\textbf{M}-dimensional} conservation
laws%
\begin{equation*}
\partial _{t_{m}}\bigl(A_{m,q}(\mathbf{u},\mathbf{u}_{x},\ldots)\bigr)=0,%
\text{ \ }m=1,\ldots,M,
\end{equation*}%
where summation is supposed over $m$ and we identify $x=t_{1}$, $t=t_{2}$, $%
y=t_{3}$ etc.\footnote{%
Simultaneously, in the three-dimensional case: $A_{q}\equiv
A_{3,q},B_{q}\equiv A_{2,q},C_{q}\equiv A_{1,q}$; in the four-dimensional
case $A_{q}=A_{4,q},B_{q}\equiv A_{3,q},C_{q}\equiv A_{2,q},D_{q}\equiv
A_{1,q}$, etc.}

Our claim is that \textit{this is a universal property of two-dimensional
integrable systems}. Without loss of generality and for simplicity we
restrict our consideration on two most remarkable cases: the dispersionless
limit of the Kadomtsev--Petviashvili equation (also known as the
Khokhlov--Zabolotzkaya equation in nonlinear acoustics or the
Lin--Reissner--Tsien equation in aerodynamics) and the Mikhal\"{e}v equation
\cite{Mikh}. These three-dimensional equations are connected with
hydrodynamic chains and possess infinitely many two-dimensional reductions.
Any of these reductions is an integrable system. Pairs of such
two-dimensional integrable systems have infinitely many local
three-dimensional conservation laws. Corresponding three-dimensional
integrable quasilinear equations possess infinitely many quasilocal
three-dimensional conservation laws. Our approach is based on construction
of two-parametric families of generating equations of quasilocal
three-dimensional conservation laws. Expansion of these generating equations
with respect to these parameters allows to extract all (a finite number)
local three-dimensional conservation laws and infinitely many quasilocal
three-dimensional conservation laws. Of course, this construction is valid
for any number of dimensions and for any integrable systems.

\subsection{Example: The Korteweg de Vries Equation}

\label{sec:KdV}

The Korteweg de Vries equation ($\epsilon $ is an arbitrary constant)%
\begin{equation*}
u_{t}=\left( -\frac{3}{2}u^{2}+\epsilon ^{2}u_{xx}\right) _{x}
\end{equation*}%
follows from the compatibility condition $\bigl(\psi _{xx}\bigr)_{t}=\bigl(%
\psi _{t}\bigr)_{xx}$, where%
\begin{equation}
4\epsilon ^{2}\psi _{xx}=(\lambda +2u)\psi ,\quad \psi _{t}=(\lambda -u)\psi
_{x}+\frac{1}{2}u_{x}\psi .  \label{2}
\end{equation}%
Introducing the function $p(\lambda )=\bigl(\psi _{1}\psi _{2}\bigr)^{-1}$,
where $\psi _{1}$ and $\psi _{2}$ are two functionally independent solutions
of the first equation in (\ref{2}), the second equation in (\ref{2})
transforms to the conservative form%
\begin{equation}
p_{t}=\Bigl((\lambda -u)p\Bigr)_{x},  \label{3}
\end{equation}%
while the first equation in (\ref{2}) becomes%
\begin{equation}
\epsilon ^{2}\bigl(p^{-1}(\lambda )\bigr)_{xxx}=(\lambda +2u)\bigl(%
p^{-1}(\lambda )\bigr)_{x}+p^{-1}(\lambda )u_{x}.  \label{4}
\end{equation}%
If $\lambda \rightarrow \infty $, one can choose solutions $\psi _{1}$ and $%
\psi _{2}$ with the following asymptotic behaviour%
\begin{equation*}
\psi _{1}\rightarrow \exp \left( \frac{\sqrt{\lambda }}{2\epsilon }x\right) ,%
\text{ \ }\psi _{2}\rightarrow \exp \left( -\frac{\sqrt{\lambda }}{2\epsilon
}x\right) .
\end{equation*}%
This means that the function $p(\lambda )=\bigl(\psi _{1}\psi _{2}\bigr)%
^{-1} $ has the following expansion%
\begin{equation*}
p=1+\frac{\sigma _{1}}{\lambda }+\frac{\sigma _{2}}{\lambda ^{2}}+\frac{%
\sigma _{3}}{\lambda ^{3}}+\ldots
\end{equation*}%
Substitution this series into (\ref{3}) leads to infinitely many local
\textit{two-dimensional} conservation laws%
\begin{equation*}
\bigl(\sigma _{k}\bigr)_{t}=\Bigl(\sigma _{k+1}-\sigma _{1}\sigma _{k}\Bigr)%
_{x},
\end{equation*}%
where $u=\sigma _{1}$ and all conservation law densities $\sigma
_{k}(u,u_{x},u_{xx},...)$ can be found from (\ref{4}).

Now we consider simultaneously the Korteweg de Vries equation%
\begin{equation*}
u_{t}=\left( -\frac{3}{2}u^{2}+\epsilon ^{2}u_{xx}\right) _{x}
\end{equation*}%
determined by the compatibility condition $\bigl(\psi _{xx}\bigr)_{t}=\bigl(%
\psi _{t}\bigr)_{xx}$, where%
\begin{equation*}
4\epsilon ^{2}\psi _{xx}=(\lambda +2u)\psi ,\quad \psi _{t}=(\lambda -u)\psi
_{x}+\frac{1}{2}u_{x}\psi ,
\end{equation*}%
and its first commuting flow%
\begin{equation*}
u_{y}=\left( \frac{5}{2}u^{3}-\frac{5}{2}\epsilon ^{2}u_{x}^{2}-5\epsilon
^{2}uu_{xx}+\epsilon ^{4}u_{xxxx}\right) _{x}
\end{equation*}%
determined by the compatibility condition $\bigl(\psi _{xx}\bigr)_{y}=\bigl(%
\psi _{y}\bigr)_{xx}$, where%
\begin{equation*}
4\epsilon ^{2}\psi _{xx}=(\lambda +2u)\psi ,\quad \psi _{y}=(\lambda
^{2}-\lambda u-v)\psi _{x}+\frac{1}{2}(\lambda u_{x}+v_{x})\psi .
\end{equation*}%
Here%
\begin{equation*}
v=-\frac{3}{2}u^{2}+\epsilon ^{2}u_{xx}.
\end{equation*}%
Again introducing the function $p(\lambda )=\bigl(\psi _{1}\psi _{2}\bigr)%
^{-1}$, where $\psi _{1}$ and $\psi _{2}$ are two functionally independent
solutions of the first equation in (\ref{2}), linear equations%
\begin{equation*}
\psi _{t}=(\lambda -u)\psi _{x}+\frac{1}{2}u_{x}\psi ,\text{ \ }\psi
_{y}=(\lambda ^{2}-\lambda u-v)\psi _{x}+\frac{1}{2}(\lambda u_{x}+v_{x})\psi
\end{equation*}%
take the conservative form%
\begin{equation*}
p_{t}=\Bigl((\lambda -u)p\Bigr)_{x},\text{ \ }p_{y}=\Bigl((\lambda
^{2}-\lambda u-v)p\Bigr)_{x}.
\end{equation*}%
Substitution the expansion%
\begin{equation*}
p=1+\frac{\sigma _{1}}{\lambda }+\frac{\sigma _{2}}{\lambda ^{2}}+\frac{%
\sigma _{3}}{\lambda ^{3}}+\ldots
\end{equation*}%
into generating equations of local \textit{two-dimensional} conservation laws%
\begin{equation*}
p_{t}=\Bigl((\lambda -u)p\Bigr)_{x},\text{ \ }p_{y}=\Bigl((\lambda
^{2}-\lambda u-v)p\Bigr)_{x}
\end{equation*}%
leads to infinitely many local \textit{two-dimensional} conservation laws%
\begin{equation*}
\bigl(\sigma _{k}\bigr)_{t}=\Bigl(\sigma _{k+1}-\sigma _{1}\sigma _{k}\Bigr)%
_{x},\quad \bigl(\sigma _{k}\bigr)_{y}=\Bigl(\sigma _{k+2}-\sigma _{1}\sigma
_{k+1}+\bigl(\sigma _{1}^{2}-\sigma _{2}\bigr)\sigma _{k}\Bigr)_{x},
\end{equation*}%
where $u=\sigma _{1}$ and $v=\sigma _{2}-\sigma _{1}^{2}$.

\textbf{Theorem}: \textit{The Korteweg de Vries pair of commuting flows}%
\begin{equation*}
u_{t}=\left( -\frac{3}{2}u^{2}+\epsilon ^{2}u_{xx}\right) _{x},\text{ \ }%
u_{y}=\left( \frac{5}{2}u^{3}-\frac{5}{2}\epsilon ^{2}u_{x}^{2}-5\epsilon
^{2}uu_{xx}+\epsilon ^{4}u_{xxxx}\right) _{x}
\end{equation*}%
\textit{possesses infinitely many local \textbf{three-dimensional}
conservation laws}%
\begin{equation*}
\bigl(A_{k}(\sigma _{1},\ldots ,\sigma _{k})\bigr)_{y}+\bigl(B_{k}(\sigma
_{1},\ldots ,\sigma _{k+1})\bigr)_{t}+\bigl(C_{k}(\sigma _{1},\ldots ,\sigma
_{k+2})\bigr)_{x}=0.
\end{equation*}

Indeed, below we prove that the generating equation of these local \textit{%
three-dimensional} conservation laws is%
\begin{equation*}
\bigl(p(\lambda )p(\zeta )\bigr)_{y}+\Bigl(\bigl(\sigma _{1}-\lambda -\zeta %
\bigr)p(\lambda )p(\zeta )\Bigr)_{t}+\Bigl(\bigl(\sigma _{2}-(\lambda +\zeta
)\sigma _{1}+\lambda \zeta \bigr)p(\lambda )p(\zeta )\Bigr)_{x}=0.
\end{equation*}%
Substitution the expansion ($\lambda \rightarrow \infty $)%
\begin{equation*}
p(\lambda )=1+\frac{\sigma _{1}}{\lambda }+\frac{\sigma _{2}}{\lambda ^{2}}+%
\frac{\sigma _{3}}{\lambda ^{3}}+\ldots
\end{equation*}%
yields infinitely many local \textit{three-dimensional} conservation laws%
\begin{gather*}
\bigl(\sigma _{1}^{2}\bigr)_{y}+\bigl(\sigma _{1}^{3}-2\sigma _{1}\sigma _{2}%
\bigr)_{t}+\bigl(\sigma _{2}^{2}-\sigma _{1}^{2}\sigma _{2}\bigr)_{x}=0, \\
\bigl(\sigma _{1}\sigma _{2}\bigr)_{y}+\bigl(\sigma _{1}^{2}\sigma
_{2}-\sigma _{1}\sigma _{3}-\sigma _{2}^{2}\bigr)_{t}+\bigl(\sigma
_{2}\sigma _{3}-\sigma _{1}^{2}\sigma _{3}\bigr)_{x}=0, \\
\bigl(\sigma _{2}^{2}\bigr)_{y}+\bigl(\sigma _{1}\sigma _{2}^{2}-2\sigma
_{2}\sigma _{3}\bigr)_{t}+\bigl(\sigma _{2}^{3}-2\sigma _{1}\sigma
_{2}\sigma _{3}+\sigma _{3}^{2}\bigr)_{x}=0, \\
\bigl(\sigma _{1}\sigma _{3}\bigr)_{y}+\bigl(\sigma _{1}^{2}\sigma
_{3}-\sigma _{1}\sigma _{4}-\sigma _{2}\sigma _{3}\bigr)_{t}+\bigl(\sigma
_{2}\sigma _{4}-\sigma _{1}^{2}\sigma _{4}\bigr)_{x}=0, \\
\ldots \ldots ,
\end{gather*}%
where $\sigma _{k}$ determined recursively from~\eqref{4} have the following
form
\begin{equation*}
\sigma _{1}=u,\quad \sigma _{2}=u^{2}+v,\quad \sigma _{3}=\epsilon
^{2}v_{xx}+\frac{\epsilon ^{2}}{2}u_{x}^{2}+\frac{1}{2}u^{3},\quad \ldots
\end{equation*}

The structure of the paper is as follows. In Section~\ref{sec:DKP} we
investigate the dispersionless limit of the Kadomtsev--Petviashvili
equation, the Benney pair of commuting hydrodynamic chains and their
hydrodynamic reductions. We found local three-dimensional conservation laws
for these integrable systems and infinitely many quasilocal
three-dimensional conservation laws for the dispersionless limit of the
Kadomtsev--Petviashvili equation. In Subsection \ref{subsec:higher} we
rewrite all local three-dimensional conservation laws for the dispersionless
limit of the Kadomtsev--Petviashvili equation as three-dimensional
hydrodynamic conservation laws for the Benney pair of commuting hydrodynamic
chains. In Subsection \ref{subsec:tri} we construct a generating equation of
three-dimensional hydrodynamic conservation laws for the Benney pair of
commuting hydrodynamic chains. We show that these conservation laws
simultaneously are quasilocal three-dimensional conservation laws for the
dispersionless limit of the Kadomtsev--Petviashvili equation. In Subsection %
\ref{subsec:hydro} we proved that $N$ component hydrodynamic reductions of
the Benney pair of commuting flows (and simultaneously of the dispersionless
limit of the Kadomtsev--Petviashvili equation) possess $2N$ infinite series
of three-dimensional hydrodynamic conservation laws. In Subsection \ref%
{subsec:four} we demonstrate the dispersionless limit of the
Kadomtsev--Petviashvili equation and its first commuting flow have
infinitely many quasilocal four-dimensional conservation laws. In Section %
\ref{sec:Mikh} we construct a generating equation of quasilocal
three-dimensional conservation laws for the Mikhal\"{e}v equation. In
Subsection \ref{subsec:hydrored} we present a family of three-dimensional
hydrodynamic conservation laws parameterised by $2N$ arbitrary functions of
a single variable for the linearly-degenerate hydrodynamic reductions. In
Subsection \ref{subsec:finite} we show that $M$ commuting finite-dimensional
systems have $M$-dimensional hydrodynamic conservation laws. In Subsection %
\ref{subsec:fourdim} we briefly discuss construction of four-dimensional
hydrodynamic conservation laws for the Mikhal\"{e}v triple of hydrodynamic
chains. In the Appendix \ref{sec:apend} we carefully derive the
over-determined system in involution, whose solution determines the
generating function of three-dimensional hydrodynamic conservation law
densities for the Benney pair of commuting hydrodynamic chains. Finally in
Conclusion \ref{sec:conc} we discuss open problems.

\section{The Dispersionless Limit of the Kadomtsev--Petviashvili Equation}

\label{sec:DKP}

The remarkable dispersionless Kadomtsev--Petviashvili equation (see, for
instance, \cite{Krich}) can be written in three distinguish forms
\begin{equation*}
u_{tt}=\bigl(u_{y}-uu_{x}\bigr)_{x},\text{ \ }\psi _{tt}=\psi _{xy}-\psi
_{x}\psi _{xx},\text{ \ }\Omega _{tt}=\Omega _{xy}-\frac{1}{2}\Omega
_{xx}^{2},
\end{equation*}%
where $u=\psi _{x}=\Omega _{xx}$. Introducing the auxiliary field variable $%
v $ such that $u_{t}=v_{x}$, the first quasilinear equation of second order
becomes the quasilinear system of two equations%
\begin{equation*}
u_{t}=v_{x},\text{ \ }v_{t}=u_{y}-uu_{x},
\end{equation*}
which possesses \textbf{\textit{three}} hydrodynamic conservation laws (see
\cite{FerKar})
\begin{equation}
\left( \frac{u^{2}}{2}\right) _{y}-\bigl(uv\bigr)_{t}+\left( \frac{v^{2}}{2}-%
\frac{u^{3}}{3}\right) _{x}=0,\text{ \ \ }u_{y}-v_{t}-\left( \frac{u^{2}}{2}%
\right) _{x}=0,\quad u_{t}=v_{x}.  \label{a}
\end{equation}%
The dispersionless Kadomtsev--Petviashvili equation written in the potential
form%
\begin{equation*}
\psi _{tt}=\psi _{xy}-\psi _{x}\psi _{xx}
\end{equation*}%
has \textbf{\textit{four}} local conservation laws (see \cite{BFT})%
\begin{eqnarray}
\bigl(\psi _{x}\bigr)_{y}-\bigl(\psi _{t}\bigr)_{t}-\left( \frac{\psi
_{x}^{2}}{2}\right) _{x} &=&0,  \label{h} \\
\left( \frac{1}{2}\psi _{x}^{2}\right) _{y}-\bigl(\psi _{x}\psi _{t}\bigr)%
_{t}+\left( \frac{1}{2}\psi _{t}^{2}-\frac{1}{3}\psi _{x}^{3}\right) _{x}
&=&0,  \label{e} \\
\bigl(\psi _{x}\psi _{t}\bigr)_{y}+\left( -\psi _{t}^{2}-\psi _{x}\psi _{y}+%
\frac{1}{3}\psi _{x}^{3}\right) _{t}+\left( \psi _{t}\psi _{y}-\psi
_{x}^{2}\psi _{t}\right) _{x} &=&0,  \label{f} \\
\left( \frac{1}{2}\psi _{t}^{2}+\frac{1}{6}\psi _{x}^{3}\right) _{y}-\bigl(%
\psi _{y}\psi _{t}\bigr)_{t}+\left( \frac{1}{2}\psi _{y}^{2}-\frac{1}{2}\psi
_{x}^{2}\psi _{y}\right) _{x} &=&0,  \label{g}
\end{eqnarray}%
whose densities and fluxes depend on $\psi _{x},\psi _{t},\psi _{y}$ only.
Introducing hydrodynamic field variables $u=\psi _{x},v=\psi _{t}$ and $%
w=\psi _{y}$ one can see that the first two above conservation laws coincide
with two first conservation laws of (\ref{a}), while two others are new:%
\begin{eqnarray*}
\left( \frac{1}{2}v^{2}+\frac{1}{6}u^{3}\right) _{y}-\bigl(vw\bigr)%
_{t}+\left( \frac{1}{2}w^{2}-\frac{1}{2}u^{2}w\right) _{x} &=&0, \\
\bigl(uv\bigr)_{y}+\left( \frac{1}{3}u^{3}-uw-v^{2}\right) _{t}+\left(
vw-u^{2}v\right) _{x} &=&0.
\end{eqnarray*}%
The dispersionless Kadomtsev--Petviashvili equation written in the double
potential form%
\begin{equation}
\Omega _{tt}=\Omega _{xy}-\frac{1}{2}\Omega _{xx}^{2}  \label{omega}
\end{equation}%
possesses already \textbf{\textit{ten}} local conservation laws (see details
in the Appendix A)%
\begin{equation}
\bigl(A_{k}\bigr)_{y}+\bigl(B_{k}\bigr)_{t}+\bigl(C_{k}\bigr)_{x}=0,
\label{tri}
\end{equation}%
where the conservation law densities $A_{k}$ and fluxes $B_{k},C_{k}$ depend
on second order derivatives of a single function $\Omega $ only. Here we
present a list of all particular solutions $A_{k},B_{m}$ and $C_{n}$:%
\begin{align*}
A_{1}& =\Omega _{xx}^{2},\qquad \qquad \qquad \qquad \qquad \qquad & B_{1}&
=-2\Omega _{xx}\Omega _{xt}, \\
A_{2}& =\Omega _{xx}\Omega _{xt}, & B_{2}& =\frac{1}{3}\Omega
_{xx}^{3}-\Omega _{xy}\Omega _{xx}-\Omega _{xt}^{2}, \\
A_{3}& =\frac{1}{6}\Omega _{xx}^{3}+\frac{1}{2}\Omega _{xt}^{2}, & B_{3}&
=-\Omega _{xt}\Omega _{xy}, \\
A_{4}& =\Omega _{xx}\Omega _{xy}-\frac{1}{6}\Omega _{xx}^{3}, & B_{4}&
=-\Omega _{xt}\Omega _{xy}-\Omega _{xx}\Omega _{yt},\\
A_{5}& =\frac{1}{2}\Omega _{xt}\Omega _{xx}^{2}+\Omega _{xt}\Omega _{xy}, &
B_{5}& =\frac{1}{6}\Omega _{xx}^{4}-\Omega _{xt}^{2}\Omega _{xx}-\Omega
_{xy}^{2}-\Omega _{xt}\Omega _{yt}, \\
A_{6}& =\frac{1}{2}\Omega _{xx}\Omega _{xt}^{2}+\frac{1}{2}\Omega _{xy}^{2},
& B_{6}& =\frac{1}{3}\Omega _{xt}\Omega _{xx}^{3}-\Omega _{xt}\Omega
_{xy}\Omega _{xx}-\Omega _{xy}\Omega _{yt} \\
A_{7}& =\frac{1}{2}\Omega _{xt}\Omega _{xx}^{2}+\Omega _{yt}\Omega _{xx}, &
\,& -\frac{1}{3}\Omega _{xt}^{3}, \\
A_{8}& =\frac{1}{2}\Omega _{xy}\Omega _{xx}^{2}+\Omega _{xt}^{2}\Omega
_{xx}+\Omega _{xt}\Omega _{yt}, & B_{7}& =\frac{1}{6}\Omega _{xx}^{4}-\Omega
_{xt}^{2}\Omega _{xx}-\Omega _{yy}\Omega _{xx}-\Omega _{xt}\Omega _{yt}, \\
A_{9}& =\frac{1}{3}\Omega _{xt}^{3}+2\Omega _{xx}\Omega _{xy}\Omega
_{xt}+\Omega _{xy}\Omega _{yt}, & B_{8}& =\frac{2}{3}\Omega _{xt}\Omega
_{xx}^{3}-2\Omega _{xt}\Omega _{xy}\Omega _{xx}-\Omega _{xy}\Omega _{yt} \\
A_{10}& =\frac{1}{2}\Omega _{xx}^{2}\Omega _{xt}^{2}+\frac{1}{2}\Omega
_{xy}\Omega _{xt}^{2}+\frac{1}{2}\Omega _{yt}^{2} & \,& -\Omega _{xt}\Omega
_{yy}-\frac{2}{3}\Omega _{xt}^{3}, \\
& \,+\frac{1}{2}\Omega _{xx}\Omega _{xy}^{2}+\Omega _{xx}\Omega _{yt}\Omega
_{xt} & B_{9}& =\frac{2}{3}\Omega _{xy}\Omega _{xx}^{3}-2\Omega _{xt}\Omega
_{yt}\Omega _{xx}-\Omega _{xy}^{2}\Omega _{xx} \\
& \, & \,& -\Omega _{yt}^{2}-2\Omega _{xt}^{2}\Omega _{xy}-\Omega
_{xy}\Omega _{yy}, \\
& \, &\, &\, \\
C_{1}& =\Omega _{xt}^{2}-\frac{2}{3}\Omega _{xx}^{3}, & B_{10}& =\frac{1}{3}%
\Omega _{xt}\Omega _{xx}^{4}+\frac{1}{3}\Omega _{yt}\Omega _{xx}^{3}-\Omega
_{xy}\Omega _{yt}\Omega _{xx} \\
C_{2}& =\Omega _{xt}\Omega _{xy}-\Omega _{xx}^{2}\Omega _{xt}, & \,& -\Omega
_{xt}\Omega _{yy}\Omega _{xx}-\frac{2}{3}\Omega _{xt}^{3}\Omega _{xx}-\Omega
_{xt}\Omega _{xy}^{2} \\
\,C_{3}& =\frac{1}{2}\Omega _{xy}^{2}-\frac{1}{2}\Omega _{xx}^{2}\Omega
_{xy}, & \,& -\Omega _{xt}^{2}\Omega _{yt}-\Omega _{yt}\Omega _{yy}, \\
C_{4}& =\Omega _{xt}\Omega _{yt}-\frac{1}{2}\Omega _{xx}^{2}\Omega _{xy},\,
& C_{6}& =\frac{1}{15}\Omega _{xx}^{5}-\frac{1}{3}\Omega _{xy}\Omega
_{xx}^{3}-\frac{1}{2}\Omega _{xt}^{2}\Omega _{xx}^{2} \\
\,& \, & \,& +\frac{1}{2}\Omega _{yt}^{2}+\frac{1}{2}\Omega _{xt}^{2}\Omega
_{xy}, \\
C_{5}& =-\frac{2}{3}\Omega _{xt}\Omega _{xx}^{3}-\frac{1}{2}\Omega
_{yt}\Omega _{xx}^{2}\, & C_{7}& =-\frac{2}{3}\Omega _{xt}\Omega _{xx}^{3}-%
\frac{1}{2}\Omega _{yt}\Omega _{xx}^{2} \\
\,& +\frac{1}{3}\Omega _{xt}^{3}+\Omega _{xy}\Omega _{yt}, & \,& +\frac{1}{3}%
\Omega _{xt}^{3}+\Omega _{xt}\Omega _{yy}, \\
\end{align*}
\begin{align*}
C_{8}& =\frac{2}{15}\Omega _{xx}^{5}-\frac{2}{3}\Omega _{xy}\Omega
_{xx}^{3}-\Omega _{xt}^{2}\Omega _{xx}^{2}\quad \qquad & C_{10}& =\frac{1}{18%
}\Omega _{xx}^{6}-\frac{2}{3}\Omega _{xt}^{2}\Omega _{xx}^{3}-\Omega
_{xt}\Omega _{yt}\Omega _{xx}^{2} \\
\,& -\frac{1}{2}\Omega _{yy}\Omega _{xx}^{2}+\Omega _{xt}^{2}\Omega
_{xy}+\Omega _{xy}\Omega _{yy}, & \,& -\frac{1}{3}\Omega _{yy}\Omega
_{xx}^{3}-\frac{1}{2}\Omega _{xy}^{2}\Omega _{xx}^{2}+\frac{1}{6}\Omega
_{xt}^{4} \\
C_{9}& =-\frac{2}{3}\Omega _{yt}\Omega _{xx}^{3}-2\Omega _{xt}\Omega
_{xy}\Omega _{xx}^{2} & \,& +\frac{1}{6}\Omega _{xy}^{3}+\frac{1}{2}\Omega
_{yy}^{2}+\Omega _{xt}\Omega _{xy}\Omega _{yt} \\
\,& +\Omega _{xt}\Omega _{xy}^{2}+\Omega _{xt}^{2}\Omega _{yt}+\Omega
_{yt}\Omega _{yy}, & \,& +\frac{1}{2}\Omega _{xt}^{2}\Omega _{yy}.
\end{align*}%
\\[-7mm]
\textbf{Remark}: Taking into account that $\psi =\Omega _{x}$, one can see
that first above three dimensional conservation laws coincide with (\ref{e}%
), (\ref{f}), (\ref{g}), while\ differentiation (\ref{omega}) with respect
to $x$ is precisely (\ref{h}).

Below we show that the dispersionless limit of the Kadomtsev--Petviashvili
equation (\ref{omega}) possesses \textit{infinitely many local
three-dimensional conservation laws} $\bigl(A_{k}\bigr)_{y}+\bigl(B_{k}\bigr)%
_{t}+\bigl(C_{k}\bigr)_{x}=0,$ whose conservation law densities $A_{k}$ and
fluxes $B_{k},C_{k}$ depend on second order derivatives of a single function
$\Omega $ with respect to \textit{all higher time variables}.

\subsection{Higher Three-Dimensional Hydrodynamic Conservation Laws}

\label{subsec:higher}

The dispersionless Kadomtsev--Petviashvili equation is an integrable
three-dimensional quasilinear equation, which follows from the compatibility
condition $\bigl(p_{t}\,\bigr)_{y}=\bigl(p_{y}\bigr)_{t}$, where%
\begin{equation}
p_{t}=\left( \frac{p^{2}}{2}+\Omega _{xx}\right) _{x},\text{ \ \ }%
p_{y}=\left( \frac{p^{3}}{3}+\Omega _{xx}p+\Omega _{xt}\right) _{x}.
\label{Lax}
\end{equation}%
The substitution%
\begin{equation}
p=\lambda -\frac{H_{0}}{\lambda }-\frac{H_{1}}{\lambda ^{2}}-\frac{H_{2}}{%
\lambda ^{3}}-\frac{H_{3}}{\lambda ^{4}}-\frac{H_{4}}{\lambda ^{5}}-\ldots
\label{ph}
\end{equation}%
into (\ref{Lax}) leads to following consequences: $H_{0}=\Omega
_{xx},H_{1}=\Omega _{xt},H_{2}=\Omega _{xy},$%
\begin{equation}
H_{3}=\Omega _{yt}+\Omega _{xx}\Omega _{xt},\text{ \ }H_{4}=\Omega
_{yy}+\Omega _{xx}\Omega _{xy}+\Omega _{xt}^{2}-\frac{1}{3}\Omega _{xx}^{3},
\label{b}
\end{equation}%
and to two commuting infinite series of \textbf{\textit{two-dimensional}}\
conservation laws:%
\begin{equation}
\bigl(H_{0}\bigr)_{t}=\bigl(H_{1}\bigr)_{x},\text{ \ }\bigl(H_{1}\bigr)%
_{t}=\left( H_{2}-\frac{1}{2}H_{0}^{2}\right) _{x},\text{ \ }\bigl(H_{2}%
\bigr)_{t}=\bigl(H_{3}-H_{0}H_{1}\bigr)_{x},\,\,\ldots  \label{i}
\end{equation}%
\begin{equation*}
\bigl(H_{0}\bigr)_{y}=\bigl(H_{2}\bigr)_{x},\text{ \ }\bigl(H_{1}\bigr)_{y}=%
\bigl(H_{3}-H_{0}H_{1}\bigr)_{x},\text{\ }\bigl(H_{2}\bigr)_{y}=\left(
H_{4}-H_{0}H_{2}-H_{1}^{2}+\frac{1}{3}H_{0}^{3}\right) _{x},\,\,\ldots
\end{equation*}%
Taking into account (see (\ref{b}))%
\begin{gather*}
\Omega _{xx}=H_{0},\text{ \ }\Omega _{xt}=H_{1},\text{ \ }\Omega _{xy}=H_{2},%
\text{ \ }\Omega _{yt}=H_{3}-H_{0}H_{1}, \\
\Omega _{yy}=H_{4}-H_{0}H_{2}-H_{1}^{2}+\frac{1}{3}H_{0}^{3},
\end{gather*}%
\\[-9mm]
all above ten \textbf{\textit{three-dimensional}} conservation laws can be
written in the hydrodynamic type form, i.e. their densities $A_{k}$ and
fluxes $B_{k},C_{k}$ depend on $H_{m}$ only:
\begin{align*}
A_{1}& =H_{0}^{2},\qquad \qquad \qquad \qquad \qquad \qquad & B_{1}&
=-2H_{0}H_{1}, \\
A_{2}& =H_{0}H_{1}, & B_{2}& =\frac{1}{3}H_{0}^{3}-H_{2}H_{0}-H_{1}^{2},\\
A_{3}& =\frac{1}{6}H_{0}^{3}+\frac{1}{2}H_{1}^{2}, & B_{3}& =-H_{1}H_{2}, \\
A_{4}& =H_{0}H_{2}-\frac{1}{6}H_{0}^{3}, & B_{4}&
=H_{1}H_{0}^{2}-H_{3}H_{0}-H_{1}H_{2}, \\
A_{5}& =\frac{1}{2}H_{1}H_{0}^{2}+H_{1}H_{2}, & B_{5}& =\frac{H_{0}^{4}}{6}%
-H_{2}^{2}-H_{1}H_{3}, \\
A_{6}& =\frac{1}{2}H_{0}H_{1}^{2}+\frac{1}{2}H_{2}^{2}, & B_{6}& =\frac{1}{3}%
H_{1}H_{0}^{3}-\frac{H_{1}^{3}}{3}-H_{2}H_{3} \\
A_{7}& =H_{0}H_{3}-\frac{1}{2}H_{0}^{2}H_{1}, & B_{7}& =-\frac{H_{0}^{4}}{6}%
+H_{2}H_{0}^{2}+H_{1}^{2}H_{0} \\
A_{8}& =\frac{1}{2}H_{2}H_{0}^{2}+H_{1}H_{3}, & \,& -H_{4}H_{0}-H_{1}H_{3},
\\
A_{9}& =\frac{1}{3}H_{1}^{3}+H_{0}H_{2}H_{1}+H_{2}H_{3}, & B_{8}& =\frac{1}{3%
}H_{1}H_{0}^{3}+\frac{H_{1}^{3}}{3}-H_{2}H_{3}-H_{1}H_{4} \\
A_{10}& =\frac{1}{2}H_{2}H_{1}^{2}+\frac{1}{2}H_{0}H_{2}^{2}+\frac{H_{3}^{2}%
}{2} & B_{9}& =\frac{1}{3}H_{2}H_{0}^{3}+H_{1}^{2}H_{0}^{2}-H_{3}^{2} \\
& \, & \,& -H_{1}^{2}H_{2}-H_{2}H_{4}, \\
C_{1}& =H_{1}^{2}-\frac{2}{3}H_{0}^{3}, & B_{10}& =\frac{1}{3}%
H_{0}H_{1}^{3}-H_{2}^{2}H_{1} \\
C_{2}& =H_{1}H_{2}-H_{0}^{2}H_{1}, & \,& +H_{0}^{2}H_{2}H_{1}-H_{3}H_{4}, \\
\,C_{3}& =\frac{1}{2}H_{2}^{2}-\frac{1}{2}H_{0}^{2}H_{2}, & C_{6}& =\frac{%
H_{0}^{5}}{15}-\frac{1}{3}H_{2}H_{0}^{3}-H_{1}H_{3}H_{0} \\
C_{4}& =-\frac{1}{2}H_{2}H_{0}^{2}-H_{1}^{2}H_{0}+H_{1}H_{3},\, & \,& +\frac{%
H_{3}^{2}}{2}+\frac{1}{2}H_{1}^{2}H_{2}, \\
\end{align*}
\begin{align*}
C_{5}& =-\frac{1}{6}H_{1}H_{0}^{3}-\frac{1}{2}H_{3}H_{0}^{2} & C_{7}& =\frac{%
1}{6}H_{1}H_{0}^{3}-\frac{1}{2}H_{3}H_{0}^{2} \\
\,& -H_{1}H_{2}H_{0}+\frac{H_{1}^{3}}{3}+H_{2}H_{3}, & \,& -H_{1}H_{2}H_{0}-%
\frac{2H_{1}^{3}}{3}+H_{1}H_{4}, \\
C_{8}& =-\frac{H_{0}^{5}}{30}+\frac{1}{6}H_{2}H_{0}^{3}-\frac{1}{2}%
H_{1}^{2}H_{0}^{2}\qquad & C_{10}& =\frac{H_{1}^{4}}{6}+\frac{1}{2}%
H_{0}^{3}H_{1}^{2}-\frac{1}{2}H_{0}H_{2}H_{1}^{2} \\
\,& -\frac{1}{2}H_{4}H_{0}^{2}-H_{2}^{2}H_{0}+H_{2}H_{4}, & \,& -\frac{1}{2}%
H_{4}H_{1}^{2}-H_{0}^{2}H_{3}H_{1}+H_{2}H_{3}H_{1} \\
C_{9}& =\frac{1}{3}H_{1}H_{0}^{4}-\frac{1}{3}%
H_{3}H_{0}^{3}-H_{1}H_{2}H_{0}^{2} & \,& +\frac{H_{2}^{3}}{6}+\frac{H_{4}^{2}%
}{2}-H_{0}H_{2}H_{4} \\
\,& -H_{2}H_{3}H_{0}-H_{1}H_{4}H_{0}, & \,& \\
\,& +H_{1}H_{2}^{2}+H_{3}H_{4}, & \,&
\end{align*}%
So, all these three-dimensional conservation laws can be presented in the
following form%
\begin{equation*}
\bigl(A_{k}(H_{0},H_{1},\ldots ,H_{k})\bigr)_{y}+\bigl(B_{k}(H_{0},H_{1},%
\ldots ,H_{k},H_{k+1})\bigr)_{t}+\bigl(C_{k}(H_{0},H_{1},\ldots
,H_{k},H_{k+1})\bigr)_{x}=0.
\end{equation*}

It is well known, that the Benney hydrodynamic chain (see \cite{Benney},
\cite{Gibbons}, \cite{KM}, \cite{Tesh}, \cite{Zakh})%
\begin{equation*}
A_{t}^{k}=A_{x}^{k+1}+kA^{k-1}A_{x}^{0},\text{ }k=0,1,\ldots
\end{equation*}%
can be written in the conservative form (see, for instance, \cite{algebra})%
\begin{equation}
(H_{k})_{t}=\left( H_{k+1}-\frac{1}{2}\underset{m=1}{\overset{N}{\sum }}%
H_{m}H_{k-1-m}\right) _{x},  \label{firstseries}
\end{equation}%
where all two-dimensional conservation law densities $H_{k}$ are polynomials
with respect to moments $A^{m}$, and can be found by substitution of inverse
expansion%
\begin{equation}
\lambda =p+\frac{A^{0}}{p}+\frac{A^{1}}{p^{2}}+\frac{A^{2}}{p^{3}}+\ldots
\label{lambda}
\end{equation}%
into (\ref{ph}). Then $H_{0}=A^{0},\,H_{1}=A^{1},\,H_{2}=A^{2}+\bigl(A^{0}%
\bigr)^{2},\,H_{3}=A^{3}+3A^{0}A^{1},$ etc. Benney hydrodynamic chain
written in conservative form (\ref{firstseries}) is nothing but precisely
the first series of two-dimensional conservation laws in (\ref{i}). Also one
can prove in a similar way that the Benney pair of commuting hydrodynamic
chains%
\begin{equation}
A_{t}^{k}=A_{x}^{k+1}+kA^{k-1}A_{x}^{0},\text{ }k=0,1,\ldots  \label{benneya}
\end{equation}%
\begin{equation}
A_{y}^{k}=A_{x}^{k+2}+A^{0}A_{x}^{k}+kA^{k-1}A_{x}^{1}+(k+1)A^{k}A_{x}^{0},%
\text{ }k=0,1,\ldots  \label{benneyb}
\end{equation}%
can be written in conservative form (\ref{i}), i.e. the second series of
two-dimensional conservation laws in (\ref{i}) is a conservative form for
the first commuting flow (\ref{benneyb}) to Benney hydrodynamic chain (\ref%
{benneya}).

Now we prove that this Benney pair of commuting hydrodynamic chains
possesses \textit{infinitely many such three-dimensional hydrodynamic
conservation laws}.

\subsection{Three-Dimensional Hydrodynamic Conservation Laws}

\label{subsec:tri}

The above ten three-dimensional hydrodynamic conservation laws can be
separated according to the dependence on a highest density $H_{k}$, i.e. we
have: \textbf{\textit{one}} conservation law
\begin{equation*}
\bigl(A_{0}(H_{0})\bigr)_{y}+\bigl(B_{0}(H_{0},H_{1})\bigr)_{t}+\bigl(%
C_{0}(H_{0},H_{1})\bigr)_{x}=0,
\end{equation*}%
\textbf{\textit{two}} conservation laws\\[-7mm]
\begin{equation*}
\bigl(A_{1}(H_{0},H_{1})\bigr)_{y}+\bigl(B_{1}(H_{0},H_{1},H_{2})\bigr)_{t}+%
\bigl(C_{1}(H_{0},H_{1},H_{2})\bigr)_{x}=0,
\end{equation*}%
\textbf{\textit{three}} conservation laws\\[-7mm]
\begin{equation*}
\bigl(A_{2}(H_{0},H_{1},H_{2})\bigr)_{y}+\bigl(B_{2}(H_{0},H_{1},H_{2},H_{3})%
\bigr)_{t}+\bigl(C_{2}(H_{0},H_{1},H_{2},H_{3})\bigr)_{x}=0,
\end{equation*}%
\textbf{\textit{four}} conservation laws\\[-7mm]
\begin{equation*}
\bigl(A_{3}(H_{0},H_{1},H_{2},H_{3})\bigr)_{y}+\bigl(%
B_{3}(H_{0},H_{1},H_{2},H_{3},H_{4})\bigr)_{t}+\bigl(%
C_{3}(H_{0},H_{1},H_{2},H_{3},H_{4})\bigr)_{x}=0.
\end{equation*}%
This situation is very \textit{unusual} in comparison with two-dimensional
conservation laws in the theory of two-dimensional integrable hydrodynamic
chains, where just one two-dimensional conservation law%
\begin{equation*}
\bigl(A_{k}(H_{0},H_{1},\ldots ,H_{k})\bigr)_{t}+\bigl(B_{k}(H_{0},H_{1},%
\ldots ,H_{k},H_{k+1})\bigr)_{x}=0
\end{equation*}%
exists for every index $k$. The number of three-dimensional hydrodynamic
conservation laws%
\begin{equation*}
\bigl(A_{k}(H_{0},H_{1},\ldots ,H_{k})\bigr)_{y}+\bigl(B_{k}(H_{0},H_{1},%
\ldots ,H_{k},H_{k+1})\bigr)_{t}+\bigl(C_{k}(H_{0},H_{1},\ldots
,H_{k},H_{k+1})\bigr)_{x}=0
\end{equation*}%
is proportional to the index $k$. This means that the generating function of
conservation law densities $A_{k}(H_{0},H_{1},\ldots ,H_{k})$ cannot depend
on a sole parameter only in a three-dimensional case.

\textbf{Theorem}: \textit{The generating equation of three-dimensional
conservation laws}%
\begin{equation*}
\bigl(A_{k}(H_{0},H_{1},\ldots ,H_{k})\bigr)_{y}+\bigl(B_{k}(H_{0},H_{1},%
\ldots ,H_{k},H_{k+1})\bigr)_{t}+\bigl(C_{k}(H_{0},H_{1},\ldots
,H_{k},H_{k+1})\bigr)_{x}=0
\end{equation*}%
\textit{for the dispersionless Kadomtsev--Petviashvili equation depends on
two arbitrary parameters, i.e.}%
\begin{equation}
\Bigl(\bigl(p(\lambda )-p(\zeta )\bigr)^{3}\Bigr)_{y}-\Bigl(\bigl(p(\lambda
)-p(\zeta )\bigr)^{3}\bigl(p(\lambda )+p(\zeta )\bigr)\Bigr) _{t}  \label{c}
\end{equation}%
\begin{equation*}
+\left( \bigl(p(\lambda )-p(\zeta )\bigr)^{3}\biggl( \frac{1}{5}\Bigl\{%
p^{2}(\lambda )+3p(\lambda )p(\zeta )+p^{2}(\zeta )\Bigr\}+H_{0}\biggr) %
\right) _{x}=0,
\end{equation*}%
\\[-5mm]
\textit{where} $p(\lambda )\equiv p(\lambda ;H_{0},H_{1},...)$ \textit{is a
one-parametric generating function of two-dimensional conservation law
densities }$H_{k}$\textit{, and corresponding generating equations of
two-dimensional conservation laws are}%
\begin{equation*}
p_{t}=\left( \frac{p^{2}}{2}+H_{0}\right) _{x},\text{ \ }p_{y}=\left( \frac{%
p^{3}}{3}+H_{0}p+H_{1}\right) _{x}.
\end{equation*}

\textbf{Proof}: One can look for a generating equation of three-dimensional
hydrodynamic conservation laws in the following ansatz%
\begin{equation*}
\bigl(A(p,q)\bigr)_{y}+\bigl(B(p,q,H_{0})\bigr)_{t}+\bigl(C(p,q,H_{0})\bigr)%
_{x}=0,
\end{equation*}%
where we denoted $p\equiv p(\lambda )$ and $q\equiv p(\zeta )$. Then one can
express all first derivatives $C_{p}\equiv \partial C/\partial p$, $%
C_{q}\equiv \partial C/\partial q$ and $C_{0}~\equiv ~\partial C/\partial
H_{0}$ via first derivatives of the functions $A$ and $B$, i.e.%
\begin{equation*}
C_{p}=-\bigl(p^{2}+H_{0}\bigr)A_{p}-pB_{p},\text{ \ }C_{q}=-\bigl(q^{2}+H_{0}%
\bigr)A_{q}-qB_{q},\text{ \ }C_{0}=-qA_{q}-pA_{p}-B_{q}-B_{p}.
\end{equation*}%
The compatibility conditions\footnote{%
Here $\bigl(C_{p}\bigr)_{q}=C_{pq}\equiv \partial ^{2}C/\partial p\partial q$
etc.}
\begin{equation*}
\bigl(C_{p}\bigr)_{q}=\bigl(C_{q}\bigr)_{p},\quad \bigl(C_{p}\bigr)_{0}=%
\bigl(C_{0}\bigr)_{p},\quad \bigl(C_{q}\bigr)_{0}=\bigl(C_{0}\bigr)_{q}
\end{equation*}%
lead to second derivatives of the function $B$ expressed via derivatives of
the function $A$, i.e.
\begin{gather*}
B_{pp}=-2pA_{pp},\quad B_{pq}=-\bigl(p+q\bigr)A_{pq},\quad B_{qq}=-2qA_{qq},
\\
B_{0p}=-A_{pq}-A_{pp},\quad B_{0q}=-A_{qq}-A_{pq},\quad B_{00}=0.
\end{gather*}%
The compatibility conditions\footnote{%
Here $\bigl(B_{0p}\bigr)_{q}=B_{0pq}\equiv \partial ^{3}C/\partial p\partial
q\partial H_{0}$ etc.}
\begin{align*}
\bigl(B_{pp}\bigr)_{q}=\bigl(B_{pq}\bigr)_{p},\quad \bigl(B_{pp}\bigr)_{0}=%
\bigl(B_{p0}\bigr)_{p},\quad \bigl(B_{pq}\bigr)_{q}=\bigl(B_{qq}\bigr)_{p},&
\\
\bigl(B_{pq}\bigr)_{0}=\bigl(B_{p0}\bigr)_{q},\quad \bigl(B_{p0}\bigr)_{q}=%
\bigl(B_{q0}\bigr)_{p},& \\
\bigl(B_{qq}\bigr)_{0}=\bigl(B_{q0}\bigr)_{q},&
\end{align*}%
lead to the system in involution:%
\begin{equation}
A_{ppp}=-\frac{A_{pq}}{p-q},\quad A_{ppq}=\frac{A_{pq}}{p-q},\quad A_{pqq}=-%
\frac{A_{pq}}{p-q},\quad A_{qqq}=\frac{A_{pq}}{p-q}.  \label{d}
\end{equation}%
This system can be integrated immediately, i.e.%
\begin{equation*}
dA_{pq}=\frac{A_{pq}}{p-q}dp-\frac{A_{pq}}{p-q}dq=A_{pq}d\ln (p-q).
\end{equation*}%
This means that $A_{pq}=p-q$ up to an arbitrary factor. Then (\ref{d})
reduces to the form%
\begin{equation*}
A_{ppp}=-1,\text{ \ }A_{ppq}=1,\text{ \ }A_{pqq}=-1,\text{ \ }A_{qqq}=1,
\end{equation*}%
whose solution is $A=-(p-q)^{3}/6$. The factor $-1/6$ is inessential. So,
one can fix $A=(p-q)^{3}$. The functions $B$ and $C$ can be found in
quadratures. The Theorem is proved.

\textbf{Remark}: One can look for generating functions of three-dimensional
hydrodynamic conservation law densities in the infinitely many various
forms: $a_{k}(p,H_{0},H_{1},\ldots,H_{k})$, i.e. $a_{0}(p,H_{0})$, $%
a_{1}(p,H_{0},H_{1})$, $a_{2}(p,H_{0},H_{1},H_{2})$ etc. Similar
computations lead to corresponding generating equations of three-dimensional
hydrodynamic conservation laws are
\begin{gather*}
\left( \frac{1}{2}p^{2}(\lambda )\right) _{y}-\left( \frac{1}{3}%
p^{3}(\lambda )+p(\lambda )H_{0}\right) _{t}+\left( \frac{1}{2}%
H_{0}^{2}\right) _{x}=0, \\
\, \\[-7mm]
\left( -\frac{1}{6}p^{3}(\lambda )+p(\lambda )H_{0}\right) _{y}+\left( \frac{%
1}{6}p^{4}(\lambda )-p(\lambda )H_{1}-H_{0}^{2}\right) _{t}+ \\
\left( -\frac{1}{30}p^{5}(\lambda )-\frac{1}{6}p^{3}(\lambda )H_{0}+\frac{1}{%
2}p^{2}(\lambda )H_{1}-p(\lambda )H_{0}^{2}+H_{0}H_{1}\right) _{x}=0, \\
\, \\[-7mm]
\left( -\frac{1}{2}p^{2}(\lambda )H_{0}+p(\lambda )H_{1}\right) _{y}+\left(
\frac{1}{3}p^{3}(\lambda )H_{0}+p(\lambda )\Bigl(H_{0}^{2}-H_{2}\Bigr)%
-H_{0}H_{1}\right) _{t}+ \\
\left( -\frac{1}{3}p^{3}(\lambda )H_{1}+\frac{1}{2}p^{2}(\lambda
)H_{2}-p(\lambda )H_{0}H_{1}-\frac{2}{3}H_{0}^{3}+H_{0}H_{2}\right) _{x}=0,\,
\\
\ldots\ldots .
\end{gather*}%
Alternatively one can expand (\ref{c}) with respect to the parameter $\zeta $%
. Such an expansion of (\ref{c}) leads to linear combinations of above
generating equations of three-dimensional conservation laws with
two-dimensional conservation laws (see (\ref{Lax})).

On the other side, the dispersionless limit of the Kadomtsev--Petviashvili
equation belongs to the integrable hierarchy (see \cite{egor}), whose
two-dimensional conservation law densities $H_{k}$ are nothing but second
order derivatives of a sole function $\Omega $ with respect to higher time
variables $t_{k}$, where we denoted $x=t_{0},t=t_{1}$ and $y=t_{2}$. Thus,
taking into account that%
\begin{equation*}
H_{k}=\Omega _{0k}\equiv \frac{\partial ^{2}\Omega }{\partial t_{0}\partial
t_{k}},
\end{equation*}%
all three-dimensional conservation law densities and corresponding fluxes
can be expressed in terms of $\Omega _{0k}$ only.

\textbf{Definition}: \textit{We call densities and fluxes of conservation
laws for the dispersionless limit of the Kadomtsev--Petviashvili equation
\textbf{quasilocal}, when they just depend on derivatives of a single
function }$\Omega $\textit{\ with respect to higher time variables}.

So, the dispersionless limit of the Kadomtsev--Petviashvili equation (\ref%
{omega}), indeed, possesses infinitely many quasilocal three-dimensional
conservation laws $\bigl(A_{k}\bigr)_{y}+\bigl(B_{k}\bigr)_{t}+\bigl(C_{k}%
\bigr)_{x}=0$, whose densities $A_{k}$ and fluxes $B_{k},C_{k}$ depend on
second order derivatives of a single function $\Omega $.

\subsection{Hydrodynamic Reductions}

\label{subsec:hydro}

Here we consider two-dimensional hydrodynamic reductions of the
dispersionless limit of the Kadomtsev--Petviashvili equation and show that
pairs of commuting two-dimensional hydrodynamic reductions possess
infinitely many local three-dimensional hydrodynamic conservation laws.

J. Gibbons and S.P. Tsarev proved (see detail in \cite{gt}, see also \cite%
{Kodama}) that the dispersionless limit of the Kadomtsev--Petviashvili
equation has infinitely many $N$ component two-dimensional hydrodynamic
reductions
\begin{equation*}
u_{t}^{i}=\left( \frac{(u^{i})^{2}}{2}+A^{0}(\mathbf{u})\right) _{x},\text{
\ }u_{y}^{i}=\left( \frac{(u^{i})^{3}}{3}+A^{0}(\mathbf{u})u^{i}+A^{1}(%
\mathbf{u})\right) _{x},
\end{equation*}%
where the functions\footnote{%
In this Subsection we use the notation $A^{k}$ for moments following D.J.
Benney, who introduced them in his seminal paper \cite{Benney}. So we would
like emphasise the difference between moment $A^{k}$ and three-dimensional
conservation law densities $A_{k}$ in this article.} $A^{0}(\mathbf{u})$ and
$A^{1}(\mathbf{u})$ are not arbitrary and can be found by substitution the
ansatz $A^{k}(\mathbf{u})$ into Benney pair of commuting hydrodynamic chains
(\ref{benneya}), (\ref{benneyb}), simultaneously. All these functions $A^{k}(%
\mathbf{u})$ can be found in quadratures, except the function $A^{0}(\mathbf{%
u})$ satisfying the nonlinear Gibbons--Tsarev system (see \cite{maksbenney},
\cite{algebra})%
\begin{equation*}
(u^{i}-u^{k})\partial _{i}\partial _{k}A^{0}=\partial _{k}A^{0}\cdot \delta
\partial _{i}A^{0}-\partial _{i}A^{0}\cdot \partial _{k}A^{0},\quad i\neq k,
\end{equation*}%
where we denote $\partial _{k}\equiv \partial /\partial A^{k}$ and the shift
operator is $\delta =\sum\limits_{m} \partial _{m}$.

Without loss of generality we consider the so-called waterbag reduction (see
also, for instance, \cite{Bogdan}, \cite{gt})%
\begin{equation}
u_{t}^{i}=\left( \frac{(u^{i})^{2}}{2}+A^{0}(\mathbf{u})\right) _{x},\quad
u_{y}^{i}=\left( \frac{(u^{i})^{3}}{3}+A^{0}(\mathbf{u})u^{i}+A^{1}(\mathbf{u%
})\right) _{x},  \label{waterbag}
\end{equation}%
where%
\begin{equation*}
A^{k}(\mathbf{u})=\frac{1}{k+1}\underset{m=1}{\overset{N}{\sum }}\epsilon
_{m}(u^{m})^{k+1},\quad \underset{m=1}{\overset{N}{\sum }}\epsilon _{m}=0
\end{equation*}%
and $\epsilon _{k}$ are arbitrary constants. Substituting these expressions
into (\ref{lambda}), one can immediately find that the waterbag reduction is
associated with the Riemann surface%
\begin{equation}
\lambda =p-\underset{m=1}{\overset{N}{\sum }}\epsilon _{m}\ln (p-u^{m}).
\label{water}
\end{equation}%
Thus, a pair of commuting two-dimensional hydrodynamic type systems (\ref%
{waterbag}) possesses infinitely many three-dimensional hydrodynamic
conservation laws (\ref{c}) where the dependence $p(\lambda )$ is inverse to
(\ref{water}). However, in comparison with the Benney pair of commuting
hydrodynamic chains (\ref{i}) any pair of $N$ component two-dimensional
hydrodynamic reductions possesses $2N$ infinite series of three-dimensional
hydrodynamic conservation laws. For instance, the generating function of
two-dimensional hydrodynamic conservation law densities $p(\lambda )$ for
the waterbag reduction has $N$ branches: $p(\lambda )=u^{k}+\lambda
_{k}p_{1}^{(k)}+\lambda _{k}^{2}p_{2}^{(k)}+\ldots$, where $\lambda
_{k}=e^{-\lambda /\epsilon _{k}}$ is a local parameter. Then all
two-dimensional hydrodynamic conservation law densities $p_{m}^{(k)}$ can be
found directly from (\ref{water}). For example (see \cite{maksbenney}, \cite%
{algebra}),%
\begin{equation*}
p_{1}^{(k)}=e^{u^{k}/\epsilon _{k}}\underset{m\neq k}{\dprod }%
(u^{k}-u^{m})^{-\epsilon _{m}/\epsilon _{k}}.
\end{equation*}%
Thus, the substitution $p(\zeta )=u^{k}+\zeta _{k}p_{1}^{(k)}+\zeta
_{k}^{2}p_{2}^{(k)}+\ldots$ into (\ref{c}) yields $N$ infinite series of
one-parametric generating equations of three-dimensional hydrodynamic
conservation laws:%
\begin{gather*}
\Bigl(\bigl(p(\lambda )-u^{k}\bigr)^{3}\Bigr)_{y}-\Bigl(\bigl(p(\lambda
)+u^{k}\bigr)\bigl(p(\lambda )-u^{k}\bigr)^{3}\Bigr)_{t} \\
+\left( \bigl(p(\lambda )-u^{k}\bigr)^{3}\biggl(\frac{1}{5}\Bigl\{%
p^{2}(\lambda )+3u^{k}p(\lambda )+(u^{k})^{2}\Bigr\}+\underset{n=1}{\overset{%
N}{\sum }}\epsilon _{n}u^{n}\biggr)\right) _{x}=0, \\
\\[-7mm]
\Bigl(3\bigl(p(\lambda )-u^{k}\bigr)^{2}\,p_{1}^{(k)}\Bigr)_{y}-\Bigl(2\bigl(%
p(\lambda )+2u^{k}\bigr)\bigl(p(\lambda )-u^{k}\bigr)^{2}\,p_{1}^{(k)}\Bigr)%
_{t} \\
+\left( \biggl(2u^{k}p(\lambda )+(u^{k})^{2}+3\underset{n=1}{\overset{N}{%
\sum }}\epsilon _{n}u^{n}\biggr)\bigl(p(\lambda )-u^{k}\bigr)%
^{2}\,p_{1}^{(k)}\right) _{x}=0, \\
\\[-9mm]
\ldots \ldots .
\end{gather*}%
Then substitution $p(\lambda )=u^{m}+\lambda _{m}p_{1}^{(m)}+\lambda
_{m}^{2}p_{2}^{(m)}+\ldots$ into these $N$ infinite series of one-parametric
generating equations of three-dimensional hydrodynamic conservation laws
leads to $2N$ infinite series of three-dimensional hydrodynamic conservation
laws. For instance, first $N(N+1)/2$ three-dimensional hydrodynamic
conservation laws are%
\begin{equation*}
\biggl(\Bigl(p_{1}^{(k)}\Bigr)^{3}\biggr)_{y}-\biggl(2u^{k}\left(
p_{1}^{(k)}\right) ^{3}\biggr)_{t}+\biggl(\left( p_{1}^{(k)}\right) ^{3}%
\biggl((u^{k})^{2}+\underset{n=1}{\overset{N}{\sum }}\epsilon _{n}u^{n}%
\biggr)\biggr)_{x}=0,
\end{equation*}%
\begin{equation*}
\Bigl(\bigl(u^{m}-u^{k}\bigr)^{3}\Bigr)_{y}-\Bigl(\bigl(u^{m}+u^{k}\bigr)%
\bigl(u^{m}-u^{k}\bigr)^{3}\Bigr)_{t}
\end{equation*}%
\begin{equation*}
+\biggl(\bigl(u^{m}-u^{k}\bigr)^{3}\biggl(\frac{1}{5}\Bigl\{%
(u^{m})^{2}+3u^{k}u^{m}+(u^{k})^{2}\Bigr\}+\underset{n=1}{\overset{N}{\sum }}%
\epsilon _{n}u^{n}\biggr)\biggr)_{x}=0,\quad m\neq k.
\end{equation*}

\subsection{Four-Dimensional Local Conservation Laws}

\label{subsec:four}

Here we again remind that our claim is that any $M-1$ commuting \textit{%
two-dimensional} integrable systems possess infinitely many $M$-dimensional
conservation laws. The case $M=3$ already is investigated. Now we consider
the case $M=4$. Here we discuss four-dimensional local conservation laws for
the first \textbf{three} \textit{two-dimensional} commuting Benney
hydrodynamic chains and simultaneously for \textbf{two} commuting \textit{%
three-dimensional} quasilinear equations (of second and third orders,
respectively). The dispersionless limit of the Kadomtsev--Petviashvili
equation together with its first commuting flow%
\begin{equation}
\Omega _{xy}=\Omega _{tt}+\frac{1}{2}\Omega _{xx}^{2},\text{ \ }\Omega
_{xxz}=\Omega _{ttt}+\Omega _{xt}\Omega _{xxx}+2\Omega _{xx}\Omega _{xxt}
\label{dkpair}
\end{equation}%
have infinitely many \textit{four-dimensional} conservation laws $\bigl(A_{k}%
\bigr)_{z}+\bigl(B_{k}\bigr)_{y}+\bigl(C_{k}\bigr)_{t}+\bigl(D_{k}\bigr)%
_{x}=0$, whose densities $A_{k}$ and fluxes $B_{k},C_{k},D_{k}$ depend on
second derivatives of the sole function $\Omega $ with respect to all higher
time variables.

Indeed, the compatibility conditions $\bigl(p_{t}\bigr)_{y}=\bigl(p_{y}\bigr)%
_{t}$, $\bigl(p_{t}\bigr)_{z}=\bigl(p_{z}\bigr)_{t}$, $\bigl(p_{y}\bigr)_{z}=%
\bigl(p_{z}\bigr)_{y}$, where
\begin{gather*}
p_{t}=\left( \frac{p^{2}}{2}+\Omega _{xx}\right) _{x},\,p_{y}=\left( \frac{%
p^{3}}{3}+p\Omega _{xx}+\Omega _{xt}\right) _{x},\, p_{z}=\left( \frac{p^{4}%
}{4}+p^{2}\Omega _{xx}+p\Omega _{xt}+\frac{\Omega _{xx}^{2}}{2}+\Omega
_{xy}\right) _{x},
\end{gather*}%
lead to three algebraic equations connecting second order derivatives of the
function $\Omega $, i.e.%
\begin{equation}
\Omega _{tt}=\Omega _{xy}-\frac{1}{2}\Omega _{xx}^{2},\quad \Omega
_{xz}=\Omega _{yt}+\Omega _{xx}\Omega _{xt},\quad \Omega _{zt}=\Omega _{yy}+%
\frac{1}{2}\Omega _{xt}^{2}-\frac{1}{3}\Omega _{xx}^{3}.  \label{eqns}
\end{equation}%
The compatibility condition $\bigl(\Omega _{yt}\bigr)_{y}=\bigl(\Omega _{yy}%
\bigr)_{t}$ leads to (\ref{dkpair}). Then, substitution (\ref{ph}) into
above generating equations of \textit{two-dimensional} conservation laws
(cf. (\ref{Lax})) leads to the three Benney commuting hydrodynamic chains:%
\begin{gather*}
\bigl(H_{1}\bigr)_{t}=\left( -\frac{1}{2}H_{0}+H_{2}\right) _{x},\,\,\,\bigl(%
H_{1}\bigr)_{y}=\bigl(H_{3}-H_{0}H_{1}\bigr)_{x},\,\,\,\bigl(H_{1}\bigr)%
_{z}=\left( -\frac{1}{2}H_{1}^{2}-H_{0}H_{2}+H_{4}\right) _{x}, \\
\bigl(H_{2}\bigr)_{t}=\bigl(H_{3}-H_{0}H_{1}\bigr)_{x},\,\,\,\bigl(H_{2}%
\bigr)_{y}=\left( \frac{1}{3}H_{0}^{3}-H_{0}H_{2}-H_{1}^{2}+H_{4}\right)
_{x}, \\
\bigl(H_{2}\bigr)_{z}=\left(
H_{0}^{2}H_{1}-H_{0}H_{3}-2H_{1}H_{2}+H_{5}\right) _{x},\,\,\,\bigl(H_{3}%
\bigr)_{t}=\left( -\frac{1}{2}H_{1}^{2}-H_{0}H_{2}+H_{4}\right) _{x}, \\
\bigl(H_{3}\bigr)_{y}=\left(
H_{0}^{2}H_{1}-H_{0}H_{3}-2H_{1}H_{2}+H_{5}\right) _{x},\ldots,
\end{gather*}%
where we denoted $H_{0}=\Omega _{xx}$, $H_{1}=\Omega _{xt},$ $H_{2}=\Omega
_{xy}$, $H_{3}=\Omega _{xz}$ and (cf. (\ref{b}) and (\ref{eqns}))%
\begin{equation*}
H_{4}=\Omega _{yy}+\Omega _{xx}\Omega _{xy}+\Omega _{xt}^{2}-\frac{1}{3}%
\Omega _{xx}^{3},\quad H_{5}=\Omega _{yz}-\Omega _{xt}\Omega
_{xx}^{2}+\Omega _{xx}\Omega _{xz}+2\Omega _{xt}\Omega _{xy}.
\end{equation*}%
Now one can look for a generating equation of four-dimensional hydrodynamic
conservation laws $A_{z}+B_{y}+C_{t}+D_{x}=0$, where%
\begin{gather*}
A=A(p(\lambda ),p(\zeta ),p(\eta )),\text{ \ }B=B(p(\lambda ),p(\zeta
),p(\eta ),H_{0}),\text{ \ }C=C(p(\lambda ),p(\zeta ),p(\eta ),H_{0},H_{1}),
\\
D=D(p(\lambda ),p(\zeta ),p(\eta ),H_{0},H_{1}).
\end{gather*}%
As in the previous (three-dimensional) case, these functions $A,B,C,D$ can
be found explicitly. However, corresponding expressions are huge as well as
\textit{four-dimensional} conservation laws. So, we present here just the
simplest example:%
\begin{gather*}
\Bigl(4\Omega _{xt}\Omega _{xy}+5\Omega _{xz}\Omega _{xx}-2\Omega
_{xt}\Omega _{xx}^{2}\Bigr)_{z}-\left( 3\Omega _{xt}\Omega _{xz}-3\Omega
_{xx}\Omega _{xt}^{2}+\frac{1}{4}\Omega _{xx}^{4}\right) _{y} \\
-\Bigl(6\Omega _{zt}\Omega _{xt}+\Omega _{xt}^{3}+10\Omega _{xz}\Omega
_{xy}+5\Omega _{yz}\Omega _{xx}-\Omega _{xt}\Omega _{xx}^{3}\Bigr)_{t} \\
+\left( 5\Omega _{yz}\Omega _{xt}+4\Omega _{xz}^{2}+6\Omega _{zt}\Omega
_{xy}-13\Omega _{xt}\Omega _{xz}\Omega _{xx}-3\Omega _{zt}\Omega _{xx}^{2}+%
\frac{3}{2}\Omega _{xt}^{2}\Omega _{xx}^{2}+\frac{1}{5}\Omega
_{xx}^{5}\right) _{x}=0.
\end{gather*}

\section{Mikhal\"{e}v Equation}

\label{sec:Mikh}

In this Section we consider another remarkable three-dimensional quasilinear
equation of second order. The Mikhal\"{e}v equation (see \cite{Mikh})%
\begin{equation}
w_{xy}=w_{tt}+w_{x}w_{xt}-w_{t}w_{xx}  \label{mikheq}
\end{equation}%
has global solutions in comparison with the dispersionless limit of the
Kadomtsev--Petviashvili equation (see detail in \cite{FKK}, \cite{energy}).
This is a linearly degenerate three-dimensional equation, which possesses
infinitely many hydrodynamic and dispersive two-dimensional integrable
reductions (see also \cite{AdlerShabat}, \cite{mas}, \cite{ShabatJNMP}). Its
simplest two-dimensional dispersive reduction is the Korteweg de Vries pair
of commuting equations considered in the Introduction. So all multi-phase
solutions of the Korteweg de Vries commuting pair are simultaneously global
solutions of the Mikhal\"{e}v equation. Under the potential substitutions $%
u=w_{x}$ and $v=w_{t}$, Mikhal\"{e}v equation becomes the Mikhal\"{e}v system%
\begin{equation}
u_{t}=v_{x},\text{ \ \ }u_{y}=v_{t}+uv_{x}-vu_{x},  \label{mikh}
\end{equation}%
which follows from the compatibility condition $\bigl(p_{t}\,\bigr)_{y}=%
\bigl(p_{y}\bigr)_{t}$, where%
\begin{equation}
p_{t}=\Bigl((\lambda -u)p\Bigr)_{x},\text{ \ }p_{y}=\Bigl((\lambda
^{2}-\lambda u-v)p\Bigr)_{x}.  \label{uno}
\end{equation}%
Also Mikhal\"{e}v system (\ref{mikh}) can be derived from two commuting
hydrodynamic chains%
\begin{equation}
\bigl(\sigma _{k}\bigr)_{t}=\Bigl(\sigma _{k+1}-\sigma _{1}\sigma _{k}\Bigr)%
_{x},\text{ \ }\bigl(\sigma _{k}\bigr)_{y}=\Bigl(\sigma _{k+2}-\sigma
_{1}\sigma _{k+1}+\bigl(\sigma _{1}^{2}-\sigma _{2}\bigr)\sigma _{k}\Bigr)%
_{x},  \label{mihach}
\end{equation}%
which can be extracted from both generating equations of two-dimensional
conservation laws (\ref{uno}) by substitution of the expansion%
\begin{equation}
p(\lambda )=1+\frac{\sigma _{1}}{\lambda }+\frac{\sigma _{2}}{\lambda ^{2}}+%
\frac{\sigma _{3}}{\lambda ^{3}}+\ldots  \label{expand}
\end{equation}%
In this case one can see that $u=\sigma _{1}$ and $v=\sigma _{2}-\sigma
_{1}^{2}$. Then one can choose two first equations $\bigl(\sigma _{1}\bigr)%
_{t}=\bigl(\sigma _{2}-\sigma _{1}^{2}\bigr)_{x}$, $\bigl(\sigma _{2}\bigr)%
_{t}=\bigl(\sigma _{3}-\sigma _{1}\sigma _{2}\bigr)_{x}$ of the first
hydrodynamic chain and the first equation $\bigl(\sigma _{1}\bigr)_{y}=\bigl(%
\sigma _{3}-2\sigma _{1}\sigma _{2}+\sigma _{1}^{3}\bigr)_{x}$ of the second
hydrodynamic chain. Eliminating $\sigma _{3}$, finally we obtain Mikhal\"{e}%
v system (\ref{mikh}).

One can repeat all computations presented in the previous Section. So we
omit similar derivation and formulate

\textbf{Theorem}: \textit{The Mikhal\"{e}v pair of commuting hydrodynamic
chains} (\ref{mihach}) \textit{possesses infinitely many three-dimensional
hydrodynamic conservation laws}%
\begin{equation}
\bigl(A_{k}(\sigma _{1},\ldots ,\sigma _{k})\bigr)_{y}+\bigl(B_{k}(\sigma
_{1},\ldots ,\sigma _{k},\sigma _{k+1})\bigr)_{t}+\bigl(C_{k}(\sigma
_{1},\ldots ,\sigma _{k},\sigma _{k+1},\sigma _{k+2})\bigr)_{x}=0,
\label{abc}
\end{equation}

\textbf{Proof}: We are looking for a two-parametric generating equation of
three-dimensional conservation laws in the form%
\begin{equation*}
\bigl(A(p,q)\bigr)_{y}+\bigl(B(p,q,\sigma _{1})\bigr)_{t}+\bigl(C(p,q,\sigma
_{1},\sigma _{2})\bigr)_{x}=0.
\end{equation*}%
where $p\equiv p(\lambda )$ and $q\equiv p(\zeta )$. Further reasons are the
same as above. One can express the first derivatives $\partial C/\partial
p\equiv C_{p}$, $\partial C/\partial q\equiv C_{q}$, $\partial C/\partial
\sigma _{1}\equiv C_{1}$ and $\partial C/\partial \sigma _{2}\equiv C_{2}$
via first derivatives of the functions $A$ and $B$, i.e.
\begin{gather*}
C_{p}=\bigl(\sigma _{1}-\lambda \bigr)B_{p}-\bigl(\lambda ^{2}-\sigma
_{1}\lambda +\sigma _{1}^{2}-\sigma _{2}\bigr)A_{p},\quad
C_{2}=-B_{1}+pA_{p}+qA_{q}, \\
C_{1}=pB_{p}+qB_{q}+2\sigma _{1}B_{1}-\bigl(2\sigma _{1}p-\lambda p\bigr)%
A_{p}-\bigl(2\sigma _{1}q-\lambda q\bigr)A_{q}, \\
C_{q}=\bigl(\sigma _{1}-\zeta \bigr)B_{q}-\bigl(\zeta ^{2}-\sigma _{1}\zeta
+\sigma _{1}^{2}-\sigma _{2}\bigr)A_{q}.
\end{gather*}%
The compatibility conditions
\begin{align*}
\bigl(C_{p}\bigr)_{q}=\bigl(C_{q}\bigr)_{p},\quad \bigl(C_{p}\bigr)_{1}=%
\bigl(C_{1}\bigr)_{p},\quad \bigl(C_{p}\bigr)_{2}=\bigl(C_{2}\bigr)_{p},& \\
\bigl(C_{q}\bigr)_{1}=\bigl(C_{1}\bigr)_{q},\quad \bigl(C_{q}\bigr)_{2}=%
\bigl(C_{2}\bigr)_{q},& \\
\bigl(C_{1}\bigr)_{2}=\bigl(C_{2}\bigr)_{1},&
\end{align*}%
lead to equations on second derivatives of the function $B$ expressed via
derivatives of the function $A$, i.e.
\begin{gather*}
B_{pp}=\bigl(\sigma _{1}-2\lambda \bigr)A_{pp},\quad B_{pq}=\bigl(\sigma
_{1}-\lambda -\zeta \bigr)A_{pq},\quad B_{p1}=qA_{pq}+pA_{pp}, \\
B_{qq}=\bigl(\sigma _{1}-2\zeta \bigr)A_{qq},\quad
B_{q1}=qA_{qq}+pA_{pq},\quad B_{11}=0,
\end{gather*}%
Their compatibility conditions lead to the simple system in involution
\begin{equation*}
A_{ppp}=A_{ppq}=A_{pqq}=A_{qqq}=0.
\end{equation*}%
Thus we found that the generating equation of three-dimensional hydrodynamic
conservation laws for Mikhal\"{e}v pair of commuting hydrodynamic chains (%
\ref{mihach})
\begin{equation}
\bigl(p(\lambda )p(\zeta )\bigr)_{y}+\Bigl(\bigl(\sigma _{1}-\lambda -\zeta %
\bigr)p(\lambda )p(\zeta )\Bigr)_{t}+\Bigl(\bigl(\sigma _{2}-(\lambda +\zeta
)\sigma _{1}+\lambda \,\zeta \bigr)p(\lambda )p(\zeta )\Bigr)_{x}=0
\label{Mikh_gen}
\end{equation}%
also depends on two arbitrary parameters $\lambda ,\zeta $. Then
substitution (\ref{expand}) leads to infinitely many three-dimensional
hydrodynamic conservation laws (\ref{abc}). The Theorem is proved.

\textbf{Remark}: In the limit $\zeta \rightarrow \lambda $, above
two-parametric generating equation of three-dimensional hydrodynamic
conservation laws reduces to the one parametric generating equation of
three-dimensional hydrodynamic conservation laws%
\begin{equation*}
\Bigl(p^{2}(\lambda )\Bigr)_{y}+\Bigl(\bigl(\sigma _{1}-2\lambda \bigr)%
p^{2}(\lambda )\Bigr)_{t}+\Bigl(\bigl(\sigma _{2}-2\lambda \sigma
_{1}+\lambda ^{2}\bigr)p^{2}(\lambda )\Bigr)_{x}=0.
\end{equation*}%
Simultaneously (\ref{c}) becomes%
\begin{equation*}
\bigl({p^{\prime }}^{3}(\lambda )\bigr)_{y}-\Bigl(2{p^{\prime }}^{3}(\lambda
)p(\lambda )\Bigr)_{t}+\Bigl({p^{\prime }}^{3}(\lambda )\bigl(p^{2}(\lambda
)-H_{0}\bigr)\Bigr)_{x}=0.
\end{equation*}%
Of course, these one-parametric generating equations cannot produce new
three-dimensional conservation laws in comparison with two-parametric
generating equation \eqref{Mikh_gen}.

Taking into account (\ref{expand}) and expanding two-parametric generating
equation~\eqref{Mikh_gen} with respect to the parameter $\lambda $, one can
obtain infinitely many one-parametric generating equations of
three-dimensional conservation laws%
\begin{gather*}
\bigl(p(\zeta )\bigr)_{y}-\bigl(\zeta \,p(\zeta )\bigr)_{t}+\Bigl(\bigl(%
\sigma _{2}-\sigma _{1}^{2}\bigr)p(\zeta )\Bigr)_{x}=0, \\
\bigl(\sigma _{1}p(\zeta )\bigr)_{y}+\Bigl(\bigl(\sigma _{1}^{2}-\zeta
\sigma _{1}-\sigma _{2}\bigr)p(\zeta )\Bigr)_{t}+\Bigl(\bigl(\sigma
_{2}-\sigma _{1}^{2}\bigr)\zeta \,p(\zeta )\Bigr)_{x}=0, \\
\bigl(\sigma _{2}p(\zeta )\bigr)_{y}+\Bigl(\bigl(\sigma _{1}\sigma
_{2}-\zeta \sigma _{2}-\sigma _{3}\bigr)p(\zeta )\Bigr)_{t}+\Bigl(\bigl(%
\sigma _{2}^{2}+\zeta (\sigma _{3}-\sigma _{1}\sigma _{2})-\sigma _{1}\sigma
_{3}\bigr)p(\zeta )\Bigr)_{x}=0, \\
\ldots \ldots .
\end{gather*}%
Expansion of these generating equations with respect to the parameter $\zeta
$ yields corresponding three-dimensional hydrodynamic conservation laws
(here we omit linear combinations of two-dimensional hydrodynamic
conservation laws (\ref{mihach}))%
\begin{gather*}
\bigl(\sigma _{1}^{2}\bigr)_{y}+\bigl(\sigma _{1}^{3}-2\sigma _{1}\sigma _{2}%
\bigr)_{t}+\bigl(\sigma _{2}^{2}-\sigma _{1}^{2}\sigma _{2}\bigr)_{x}=0, \\
\bigl(\sigma _{1}\sigma _{2}\bigr)_{y}+\bigl(\sigma _{1}^{2}\sigma
_{2}-\sigma _{1}\sigma _{3}-\sigma _{2}^{2}\bigr)_{t}+\bigl(\sigma
_{2}\sigma _{3}-\sigma _{1}^{2}\sigma _{3}\bigr)_{x}=0, \\
\bigl(\sigma _{2}^{2}\bigr)_{y}+\bigl(\sigma _{1}\sigma _{2}^{2}-2\sigma
_{2}\sigma _{3}\bigr)_{t}+\bigl(\sigma _{2}^{3}-2\sigma _{1}\sigma
_{2}\sigma _{3}+\sigma _{3}^{2}\bigr)_{x}=0, \\
\bigl(\sigma _{1}\sigma _{3}\bigr)_{y}+\bigl(\sigma _{1}^{2}\sigma
_{3}-\sigma _{1}\sigma _{4}-\sigma _{2}\sigma _{3}\bigr)_{t}+\bigl(\sigma
_{2}\sigma _{4}-\sigma _{1}^{2}\sigma _{4}\bigr)_{x}=0, \\
\bigl(\sigma _{2}\sigma _{3}\bigr)_{y}+\bigl(\sigma _{1}\sigma _{2}\sigma
_{3}-\sigma _{2}\sigma _{4}-\sigma _{3}^{2}\bigr)_{t}+\bigl(\sigma
_{2}^{2}\sigma _{3}-\sigma _{1}\sigma _{2}\sigma _{4}-\sigma _{1}\sigma
_{3}^{2}+\sigma _{3}\sigma _{4}\bigr)_{x}=0, \\
\ldots \ldots .
\end{gather*}%
One can consider higher commuting flows (see \cite{energy}), the first
conservation law has the form (we remind that here $%
x=t_{1},t=t_{2},y=t_{3},z=t_{4}$)%
\begin{equation*}
\bigl(\sigma _{1}\bigr)_{t_{k}}=-\bigl(a_{k}\,\bigr)_{x},
\end{equation*}%
where the functions $a_{k}(\mathbf{\sigma })$ are connected with $\sigma
_{m} $ via infinitely constraints%
\begin{equation*}
\sigma _{1}+a_{1}=0,\text{ \ }a_{2}+a_{1}\sigma _{1}+\sigma _{2}=0,\text{ \ }%
a_{3}+a_{2}\sigma _{1}+a_{1}\sigma _{2}+\sigma _{3}=0,\ldots
\end{equation*}%
Introducing the potential function $w$ such that\footnote{%
We remind that the potential function $w$ depends on infinitely time
variables $t_{k}$ and Mikhal\"{e}v equation (\ref{mikheq}) is just a
simplest three-dimensional quasilinear equation belonging to the
corresponding linearly degenerate integrable hierarchy.} $\sigma _{1}=w_{x}$
and $a_{k}=-\partial _{k}w$, we come to the conclusion that all other
two-dimensional conservation law densities also can be expressed via first
order derivatives of a single function $w$, for instance:%
\begin{equation}
\sigma _{2}=w_{x}^{2}+w_{t},\text{ \ }\sigma
_{3}=w_{x}^{3}+2w_{x}w_{t}+w_{y},\text{ \ }\sigma
_{4}=w_{x}^{4}+3w_{x}^{2}w_{t}+w_{t}^{2}+2w_{x}w_{y}+w_{z},\ldots  \label{l}
\end{equation}

\textbf{Definition}: \textit{We call densities and fluxes of conservation
laws for the Mikhal\"{e}v equation \textbf{quasilocal}, when they just
depend on derivatives of a single function }$w$\textit{\ with respect to
higher time variables}.

Thus, all above three-dimensional hydrodynamic conservation laws for the
Mikhal\"{e}v pair of hydrodynamic chains simultaneously are
three-dimensional quasilocal conservation laws of the Mikhal\"{e}v equation,
whose densities and fluxes depend on first order derivatives of a sole
function $w$, for instance:%
\begin{gather*}
\bigl(w_{x}^{2}\bigr)_{y}-\bigl(w_{x}^{3}+2w_{x}w_{t}\bigr)_{t}+\bigl(%
w_{x}^{2}w_{t}+w_{t}^{2}\bigr)_{x}=0, \\
\\[-6mm]
\Bigl(w_{x}\bigl(w_{x}^{2}+w_{t}\bigr)\Bigr)_{y}-\Bigl(%
w_{x}^{4}+3w_{x}^{2}w_{t}+w_{t}^{2}+w_{x}w_{y}\Bigr)_{t}+\Bigl(w_{t}\bigl(%
w_{x}^{3}+2w_{x}w_{t}+w_{y}\bigr)\Bigr)_{x}=0, \\
\\[-5mm]
\Bigl(\bigl(w_{t}+w_{x}^{2}\bigr)^{2}\Bigr)_{y}-\Bigl(\bigl(w_{t}+w_{x}^{2}%
\bigr)\bigl(2w_{t}w_{x}+w_{x}^{3}+2w_{y}\bigr)\Bigr)_{t} \\
+\Bigl(w_{t}\bigl(w_{x}^{4}+2w_{x}w_{y}\bigr)%
+w_{t}^{3}+3w_{t}^{2}w_{x}^{2}+w_{y}^{2}\Bigr)_{x}=0, \\
\\[-5mm]
\Bigl(w_{x}\bigl(w_{x}^{3}+2w_{t}w_{x}+w_{y}\bigr)\Bigr)_{y}-\Bigl(w_{x}%
\bigl(w_{x}^{4}+2w_{x}w_{y}+w_{x}w_{z}\bigr)+w_{t}\bigl(4w_{x}^{3}+w_{y}%
\bigr)+3w_{x}w_{t}^{2}\Bigr)_{t} \\
+\Bigl(w_{t}\bigl(w_{x}^{4}+3w_{x}^{2}w_{t}+w_{t}^{2}+2w_{x}w_{y}+w_{z}\bigr)%
\Bigr)_{x}=0, \\
\\[-4mm]
\Bigl(\bigl(w_{t}+w_{x}^{2}\bigr)\bigl(w_{x}^{3}+2w_{t}w_{x}+w_{y}\bigr)%
\Bigr)_{y} \\
-\Bigl(w_{x}^{2}\bigl(w_{x}^{4}+3w_{x}w_{y}+w_{z}\bigr)+w_{t}w_{x}\bigl(%
5w_{x}^{3}+6w_{x}w_{t}+5w_{y}\bigr)+w_{t}^{3}+w_{y}^{2}+w_{t}w_{z}\Bigr)_{t}
\\
+\Bigl(w_{t}w_{x}\bigl(w_{x}^{4}+3w_{x}w_{y}+w_{z}\bigr)+2w_{t}^{2}\bigl(%
2w_{x}^{3}+w_{y}\bigr)+w_{y}\bigl(w_{y}w_{x}+w_{z}\bigr)+3w_{t}^{3}w_{x}%
\Bigr)_{x}=0, \\
\ldots \ldots .
\end{gather*}

\subsection{Hydrodynamic Reductions}

\label{subsec:hydrored}

A wide class of two-dimensional hydrodynamic reductions of Mikhal\"{e}v
system (\ref{mikh}) is the so called \textquotedblleft $\epsilon $%
-systems\textquotedblright :%
\begin{equation}
r_{t}^{k}=\Bigl(r^{k}-u\Bigr)r_{x}^{k},\text{ \ }r_{y}^{k}=\Bigl(%
(r^{k})^{2}-ur^{k}-v\Bigr)r_{x}^{k},  \label{eps}
\end{equation}%
where ($\epsilon _{k}$ are arbitrary constants)%
\begin{equation*}
u=\underset{m=1}{\overset{N}{\sum }}\epsilon _{m}r^{m},\text{ \ \ }v=\frac{1%
}{2}\underset{m=1}{\overset{N}{\sum }}\epsilon _{m}(r^{m})^{2}-\frac{1}{2}%
\left( \underset{m=1}{\overset{N}{\sum }}\epsilon _{m}r^{m}\right) ^{2}.
\end{equation*}%
The generating function of two-dimensional conservation law densities $%
p(\lambda )\equiv p(\lambda ,\mathbf{r}(x,t))$ can be found in quadratures
directly from (\ref{uno}), i.e. (see \cite{chains})%
\begin{equation*}
p(\lambda )=\underset{m=1}{\overset{N}{\dprod }}(\lambda -r^{m})^{-\epsilon
_{m}}.
\end{equation*}

Now we consider the simplest case, i.e. all parameters $\epsilon _{k}=1$.
The corresponding pair of linearly degenerate commuting hydrodynamic type
systems (see (\ref{eps})) possesses generating equation of three-dimensional
hydrodynamic conservation laws (\ref{Mikh_gen}). Then we can integrate this
generating equation with respect to parameters $\lambda $ and $\zeta $
surrounding all simple poles $\lambda =r^{k}$, $\zeta =r^{m}$ and with
arbitrary functions $R(\lambda )$, $L(\zeta )$, i.e. generating equation (%
\ref{Mikh_gen}) takes the form $A_{y}+B_{t}+C_{x}=0$, where%
\begin{equation*}
A=\doint \doint R(\lambda )L(\zeta )p(\lambda )p(\zeta )d\lambda d\zeta ,%
\text{ \ }B=\doint \doint R(\lambda )L(\zeta )\bigl(u-\lambda -\zeta \bigr)%
p(\lambda )p(\zeta )d\lambda d\zeta ,
\end{equation*}%
\begin{equation*}
C=\doint \doint R(\lambda )L(\zeta )\bigl(v+u^{2}-(\lambda +\zeta )u+\lambda
\zeta \bigr)p(\lambda )p(\zeta )d\lambda d\zeta .
\end{equation*}%
These integrals can be found in explicit form:\\[-2mm]
\begin{equation*}
A=\underset{k=1}{\overset{N}{\sum }}\frac{R_{k}(r^{k})}{\underset{p\neq k}{%
\dprod }(r^{k}-r^{p})}\cdot \underset{m=1}{\overset{N}{\sum }}\frac{%
L_{m}(r^{m})}{\underset{q\neq m}{\dprod }(r^{m}-r^{q})},
\end{equation*}%
\\[-8mm]
\begin{equation*}
B=u\underset{k=1}{\overset{N}{\sum }}\frac{R_{k}(r^{k})}{\underset{p\neq k}{%
\dprod }(r^{k}-r^{p})}\cdot \underset{m=1}{\overset{N}{\sum }}\frac{%
L_{m}(r^{m})}{\underset{q\neq m}{\dprod }(r^{m}-r^{q})}
\end{equation*}%
\begin{equation*}
-\underset{k=1}{\overset{N}{\sum }}\frac{r^{k}R_{k}(r^{k})}{\underset{p\neq k%
}{\dprod }(r^{k}-r^{p})}\cdot \underset{m=1}{\overset{N}{\sum }}\frac{%
L_{m}(r^{m})}{\underset{q\neq m}{\dprod }(r^{m}-r^{q})}-\underset{k=1}{%
\overset{N}{\sum }}\frac{R_{k}(r^{k})}{\underset{p\neq k}{\dprod }%
(r^{k}-r^{p})}\cdot \underset{m=1}{\overset{N}{\sum }}\frac{r^{m}L_{m}(r^{m})%
}{\underset{q\neq m}{\dprod }(r^{m}-r^{q})},
\end{equation*}%
\\[-4mm]
\begin{equation*}
C=(v+u^{2})\underset{k=1}{\overset{N}{\sum }}\frac{R_{k}(r^{k})}{\underset{%
p\neq k}{\dprod }(r^{k}-r^{p})}\cdot \underset{m=1}{\overset{N}{\sum }}\frac{%
L_{m}(r^{m})}{\underset{q\neq m}{\dprod }(r^{m}-r^{q})}-u\underset{k=1}{%
\overset{N}{\sum }}\frac{r^{k}R_{k}(r^{k})}{\underset{p\neq k}{\dprod }%
(r^{k}-r^{p})}\cdot \underset{m=1}{\overset{N}{\sum }}\frac{L_{m}(r^{m})}{%
\underset{q\neq m}{\dprod }(r^{m}-r^{q})}
\end{equation*}%
\begin{equation*}
-u\underset{k=1}{\overset{N}{\sum }}\frac{R_{k}(r^{k})}{\underset{p\neq k}{%
\dprod }(r^{k}-r^{p})}\cdot \underset{m=1}{\overset{N}{\sum }}\frac{%
r^{m}L_{m}(r^{m})}{\underset{q\neq m}{\dprod }(r^{m}-r^{q})}+\underset{k=1}{%
\overset{N}{\sum }}\frac{r^{k}R_{k}(r^{k})}{\underset{p\neq k}{\dprod }%
(r^{k}-r^{p})}\cdot \underset{m=1}{\overset{N}{\sum }}\frac{r^{m}L_{m}(r^{m})%
}{\underset{q\neq m}{\dprod }(r^{m}-r^{q})},
\end{equation*}%
where $R_{k}(r^{k})$ and $L_{m}(r^{m})$ are arbitrary analytic functions.
Thus, we found three-dimensional hydrodynamic conservation laws depended on $%
2N$ arbitrary functions of a single variable for pair of commuting
hydrodynamic type systems (\ref{eps}).

\subsection{Finite-Dimensional Reductions}

\label{subsec:finite}

Linearly degenerate hydrodynamic type systems (see (\ref{eps}), $\epsilon
_{k}=1$)%
\begin{equation}
r_{t}^{k}=\Bigl(r^{k}-u\Bigr)r_{x}^{k},\text{ \ }r_{y}^{k}=\Bigl(%
(r^{k})^{2}-ur^{k}-v\Bigr)r_{x}^{k},  \label{lindeg}
\end{equation}%
where
\begin{equation*}
u=\underset{m=1}{\overset{N}{\sum }}r^{m},\text{ \ \ }v=\frac{1}{2}\underset{%
m=1}{\overset{N}{\sum }}(r^{m})^{2}-\frac{1}{2}\left( \underset{m=1}{\overset%
{N}{\sum }}r^{m}\right) ^{2}
\end{equation*}\\[-3mm]
have the general solution (see \cite{ferlin}):\\[-3mm]
\begin{gather*}
x=\underset{m=1}{\overset{N}{\sum }}\,\overset{r^{m}}{\int }\frac{\xi
^{N-1}d\xi }{S_{m}(\xi )},\text{ \ }t=\underset{m=1}{\overset{N}{\sum }}\,%
\overset{r^{m}}{\int }\frac{\xi ^{N-2}d\xi }{S_{m}(\xi )},\text{ \ }y=%
\underset{m=1}{\overset{N}{\sum }}\,\overset{r^{m}}{\int }\frac{\xi
^{N-3}d\xi }{S_{m}(\xi )}, \\
0=\underset{m=1}{\overset{N}{\sum }}\,\overset{r^{m}}{\int }\frac{\xi
^{k}d\xi }{S_{m}(\xi )},\text{ \ }k=0,\ldots,N-4,
\end{gather*}%
\\[-5mm]
where $S_{k}(r^{k})$ are arbitrary functions. Indeed, one can compute first
order derivatives $\partial x/\partial r^{k},\partial t/\partial
r^{m},\partial y/\partial r^{n}$. Inverse expressions%
\begin{equation}
r_{x}^{k}=\frac{S_{k}(r^{k})}{\underset{p\neq k}{\dprod }(r^{k}-r^{p})},%
\text{ \ }r_{t}^{k}=\Bigl(r^{k}-u\Bigr)\frac{S_{k}(r^{k})}{\underset{p\neq k}%
{\dprod }(r^{k}-r^{p})},\text{ }r_{y}^{k}=\Bigl((r^{k})^{2}-ur^{k}-v\Bigr)%
\frac{S_{k}(r^{k})}{\underset{p\neq k}{\dprod }(r^{k}-r^{p})}  \label{j}
\end{equation}%
one can substitute back into (\ref{lindeg}) and obtain identities.

In the particular case ($E_{k}$ are arbitrary parameters)%
\begin{equation*}
S_{k}(r^{k})=\sqrt{P_{2N+1}(r^{k})}=\sqrt{\underset{m=1}{\overset{2N+1}{%
\dprod }}(r^{k}-E_{m})},
\end{equation*}%
three commuting finite-dimensional systems (\ref{j}) determine the $N$-phase
solution of the Korteweg de Vries equation (see again \cite{ferlin}, and
formulae 2.3.12, 2.4.3 in \cite{DMN}, the first hydrodynamic type system in (%
\ref{lindeg}) can be found in \cite{DMN} at page 139).

So, the \textit{pair} of commuting finite-dimensional systems (see (\ref{j}))%
\begin{equation*}
r_{x}^{k}=\frac{S_{k}(r^{k})}{\underset{p\neq k}{\dprod }(r^{k}-r^{p})},%
\text{ \ }r_{t}^{k}=\Bigl(r^{k}-u\Bigr)\frac{S_{k}(r^{k})}{\underset{p\neq k}%
{\dprod }(r^{k}-r^{p})}
\end{equation*}%
has infinitely many \textit{two-dimensional} conservation laws (see \cite%
{makslin}; $T_{k}(r^{k})$ are arbitrary functions)\\[-4mm]
\begin{equation*}
\left( \underset{k=1}{\overset{N}{\sum }}\frac{T_{k}(r^{k})}{\underset{p\neq
k}{\dprod }(r^{k}-r^{p})}\right) _{t}=\left( \underset{k=1}{\overset{N}{\sum
}}\frac{(r^{k}-u)T_{k}(r^{k})}{\underset{p\neq k}{\dprod }(r^{k}-r^{p})}%
\right) _{x},
\end{equation*}%
\newline
while the \textit{triple} of commuting finite-dimensional systems (\ref{j})
has infinitely many \textit{three-dimensional} conservation laws $%
A_{y}+B_{t}+C_{x}=0$, presented in the previous Subsection. Of course,
instead of three commuting finite-dimensional systems (\ref{j}), one can
consider any number of such commuting finite-dimensional system (including
higher commuting flows). Then corresponding multi-dimensional conservation
laws can be easily found.

\subsection{Four-Dimensional Local Conservation Laws}

\label{subsec:fourdim}

We already illustrated construction of generating equations of
three-dimensional and four-dimensional quasilocal conservation laws for the
dispersionless limit of the Kadomtsev--Petviashvili hierarchy. In this
Section we constructed a generating equation of three-dimensional quasilocal
conservation laws for Mikhal\"{e}v equation (\ref{mikheq}). So, our approach
allows to construct generating equations of multi-dimensional quasilocal
conservation laws for any integrable three-dimensional hierarchy. Even if
such a hierarchy is not explicitly written. For instance, now we consider
the dispersionless Lax \textquotedblleft triple\textquotedblright
\begin{equation}
p_{t}=\Bigl((\lambda -u)p\Bigr)_{x},\text{ \ }p_{y}=\Bigl((\lambda
^{2}-\lambda u-v)p\Bigr)_{x},\text{ \ }p_{z}=\Bigl((\lambda ^{3}-\lambda
^{2}u-\lambda v-s)p\Bigr)_{x},  \label{k}
\end{equation}%
where $u=\sigma _{1}$, $v=\sigma _{2}-\sigma _{1}^{2}$ and $s=\sigma
_{1}^{3}-2\sigma _{1}\sigma _{2}+\sigma _{3}$. One can look for a generating
equation of four-dimensional conservation laws $A_{z}+B_{y}+C_{t}+D_{x}=0$
in the ansatz
\begin{equation*}
A=A(p(\lambda ),p(\zeta ),p(\eta )),\,\,B=B(p(\lambda ),p(\zeta ),p(\eta
),\sigma _{1}),\,\,C=C(p(\lambda ),p(\zeta ),p(\eta ),\sigma _{1},\sigma
_{2}),
\end{equation*}%
\begin{equation*}
D=D(p(\lambda ),p(\zeta ),p(\eta ),\sigma _{1},\sigma _{2},\sigma _{3}).
\end{equation*}%
As in the three-dimensional case, these functions can be calculated
explicitly. However, unlike to the case of the dispersionless limit of the
Kadomtsev--Petviashvili equation, the generating equation has the fairly
simple form
\begin{gather*}
\biggl(p(\lambda )\,p(\zeta )\,p(\eta )\biggr)_{z}+\biggl(\Bigl(\sigma
_{1}-\lambda -\zeta -\eta \Bigr)p(\lambda )\,p(\zeta )\,p(\eta )\biggr)_{y}
\\
+\biggl(\Bigl(\sigma _{2}-\bigl(\lambda +\zeta +\eta \bigr)\sigma
_{1}+\lambda \,\zeta +\lambda \,\eta +\zeta \,\eta \Bigr)p(\lambda
)\,p(\zeta )\,p(\eta )\biggr)_{t} \\
+\biggl(\Bigl(\sigma _{3}-\bigl(\lambda +\zeta +\eta \bigr)\sigma _{2}+\bigl(%
\lambda \,\zeta +\lambda \,\eta +\zeta \,\eta \bigr)\sigma _{1}-\lambda
\,\zeta \,\eta \Bigr)p(\lambda )\,p(\zeta )\,p(\eta )\biggr)_{x}=0.
\end{gather*}%
The first non-trivial four-dimensional hydrodynamic conservation law has the
following form%
\begin{gather*}
\Bigl(\sigma _{1}^{3}+6\sigma _{1}\sigma _{2}+3\sigma _{3}\Bigr)_{z}+\Bigl(%
\sigma _{1}^{4}-3\sigma _{1}^{2}\sigma _{2}-15\sigma _{1}\sigma _{3}-9\bigl(%
\sigma _{2}^{2}+\sigma _{4}\bigr)\Bigr)_{y} \\
+\Bigl(9\sigma_{5}-8\sigma _{1}^{3}\sigma _{2}-9\sigma _{1}^{2}\sigma _{3}+21\sigma
_{2}\sigma _{3}+\sigma _{1}\bigl(6\sigma _{2}^{2}+9\sigma _{4}\bigr)\Bigr)_{t} \\
+\Bigl(3\sigma _{1}\sigma _{5}-3\sigma
_{6}-10\sigma _{2}^{3}+10\sigma _{1}^{3}\sigma _{3}+15\sigma
_{1}^{2}\sigma _{4}-15\sigma _{2}\sigma _{4}\Bigr)_{x}=0.
\end{gather*}%
Of course, one can substitute expansion (\ref{expand}) into (\ref{k}) to
extract first three Mikhal\"{e}v hydrodynamic chains, check the
compatibility conditions $\bigl(p_{t}\,\bigr)_{y}=\bigl(p_{y}\bigr)_{t}$, $%
\bigl(p_{t}\,\bigr)_{z}=\bigl(p_{z}\bigr)_{t}$, $\bigl(p_{z}\,\bigr)_{y}=%
\bigl(p_{y}\bigr)_{z}$, and derive a corresponding linearly-degenerate
system belonging to the Mikhal\"{e}v hierarchy. However, this linearly
degenerate system also can be obtained directly from the above
four-dimensional conservation law, if to remember that all functions $\sigma
_{k}$ can be expressed from (\ref{l}) as polynomials with respect to first
derivatives of the common function $w$.

\section{Appendix}

\label{sec:apend}

In this section we compute densities\footnote{%
The ansatz for densities and fluxes does not depend on $\Omega _{tt}$ due to
the algebraic relationship between second order derivatives in (\ref{omega}).%
} $A_{k}(\Omega _{xx},\Omega _{xt},\Omega _{xy},\Omega _{yt},\Omega _{yy})$
and corresponding fluxes $B_{k}(\Omega _{xx},\Omega _{xt},\Omega
_{xy},\Omega _{yt},\Omega _{yy})$, $C_{k}(\Omega _{xx},\Omega _{xt},\Omega
_{xy},\Omega _{yt},\Omega _{yy})$ of local three-dimensional conservation
laws (\ref{tri}) for the dispersionless limit of the Kadomtsev--Petviashvili
equation written in algebraic form (\ref{omega}):%
\begin{equation*}
\Omega _{tt}=\Omega _{xy}-\frac{1}{2}\Omega _{xx}^{2}.
\end{equation*}%
As usual we rewrite the three-dimensional conservation law $%
A_{y}+B_{t}+C_{x}=0$ in the expanded form%
\begin{align*}
& \frac{\partial A}{\partial \Omega _{xx}}\Omega _{xxy}+\frac{\partial A}{%
\partial \Omega _{xt}}\Omega _{xty}+\frac{\partial A}{\partial \Omega _{xy}}%
\Omega _{xyy}+\frac{\partial A}{\partial \Omega _{yt}}\Omega _{yyt} \\
+ &\frac{\partial A}{\partial \Omega _{yy}}\Omega _{yyy}+\frac{\partial B}{%
\partial \Omega _{xx}}\Omega _{xxt}+\frac{\partial B}{\partial \Omega _{xt}}%
\Omega _{xtt}+\frac{\partial B}{\partial \Omega _{xy}}\Omega _{xty}+\frac{%
\partial B}{\partial \Omega _{yt}}\Omega _{ytt} \\
+&\frac{\partial B}{\partial \Omega _{yy}}\Omega _{yyt}+\frac{\partial C}{%
\partial \Omega _{xx}}\Omega _{xxx}+\frac{\partial C}{\partial \Omega _{xt}}%
\Omega _{xxt}+\frac{\partial C}{\partial \Omega _{xy}}\Omega _{xxy}+\frac{%
\partial C}{\partial \Omega _{yt}}\Omega _{xyt}+\frac{\partial C}{\partial
\Omega _{yy}}\Omega _{xyy}=0.
\end{align*}%
Taking into account differential consequences of the dispersionless limit of
the Kadomtsev--Petviashvili equation
\begin{equation*}
\Omega _{xtt}=\Omega _{xxy}-\Omega _{xx}\Omega _{xxx},\text{ \ }\Omega
_{ytt}=\Omega _{xyy}-\Omega _{xx}\Omega _{xxy}
\end{equation*}%
and equating factors of all third derivatives of function $\Omega $ to zero
we obtain the system
\begin{gather}
C_{u}-uB_{v}=0,\quad C_{v}+B_{u}=0,\quad C_{w}+B_{v}-uB_{r}+A_{u}=0,  \notag
\\
C_{r}+B_{w}+A_{v}=0,\quad C_{s}+B_{r}+A_{w}=0,  \label{A:syst} \\
B_{s}+A_{r}=0,\quad A_{s}=0.  \notag
\end{gather}%
Here the following notation is used $u=\Omega _{xx},v=\Omega _{xt},w=\Omega
_{xy},r=\Omega _{yt},s=\Omega _{yy}$. The last equation in~(\ref{A:syst})
implies that $A=A(u,v,w,r)$. Now we have to check compatibility conditions
\begin{align*}
\bigl(C_{u}\bigr)_{v}=\bigl(C_{v}\bigr)_{u},\quad \bigl(C_{u}\bigr)_{w}=%
\bigl(C_{w}\bigr)_{u},\quad \bigl(C_{u}\bigr)_{r}=\bigl(C_{r}\bigr)%
_{u},\quad & \bigl(C_{u}\bigr)_{s}=\bigl(C_{s}\bigr)_{u}, \\
\bigl(C_{v}\bigr)_{w}=\bigl(C_{w}\bigr)_{v},\quad \bigl(C_{v}\bigr)_{r}=%
\bigl(C_{r}\bigr)_{v},\quad & \bigl(C_{v}\bigr)_{s}=\bigl(C_{s}\bigr)_{v}, \\
\bigl(C_{w}\bigr)_{r}=\bigl(C_{r}\bigr)_{w},\quad & \bigl(C_{w}\bigr)_{s}=%
\bigl(C_{s}\bigr)_{w}, \\
& \bigl(C_{r}\bigr)_{s}=\bigl(C_{s}\bigr)_{r}.
\end{align*}%
They imply a set of equations on second derivatives of function $B$. Taking
into account differential consequences of equation $B_{s}+A_{r}=0$ from~ (%
\ref{A:syst}), one can obtain the following system:%
\begin{align*}
B_{uu}& =2uA_{uv}, & B_{uw}& =uA_{vw}-A_{uv}+uA_{ru}, \\
B_{uv}& =B_{r}-A_{uu}+uA_{vv}, & B_{vw}& =-A_{vv}-A_{uw}+uA_{rv}, \\
B_{vv}& =-2A_{uv}, & B_{ww}& =2uA_{rw}-2A_{vw},
\end{align*}%
\begin{align*}
\,\,\quad \quad B_{ru}& =uA_{rv}-A_{uw}, & \quad \,\,B_{su}& =-A_{ru}, &
B_{rs}& =-A_{rr}, \\
B_{rv}& =-A_{vw}-A_{ru}, & B_{sv}& =-A_{rv}, & B_{ss}& =0. \\
B_{rw}& =-A_{ww}-A_{rv}+uA_{rr}, & B_{sw}& =-A_{rw}, & & \\
B_{rr}& =-2A_{rw}, & & & &
\end{align*}%
Compatibility conditions $\bigl(B_{uu}\bigr)_{v}=\bigl(B_{uv}\bigr)_{u}$
etc. lead to the \textit{linear system in involution} on third derivatives
of the function $A(u,v,w,r)$:%
\begin{align*}
\quad A_{uuu}& =u^{2}A_{rr}-uA_{ww}-uA_{rv} & A_{uuv}&
=-2uA_{rw}+A_{vw}+A_{ru} \\
& +A_{vv}-A_{uw}, & A_{uvv}& =-uA_{rr}+A_{ww}+2A_{rv}, \\
& & A_{vvv}& =2A_{rw},
\end{align*}%
\begin{align*}
\,\,\,A_{uuw}& =-uA_{rrr}+A_{rv}, & A_{ruu}& =0, & A_{rru}& =0, \\
A_{uvw}& =2A_{rw}, & A_{ruv}& =A_{rr}, & A_{rrv}& =0, \\
A_{vvw}& =A_{rr}, & A_{rvv}& =0, & A_{rrw}& =0, \\
A_{uww}& =A_{rr}, & A_{ruw}& =0, & A_{rrr}& =0, \\
A_{vww}& =0, & A_{rvw}& =0, & & \\
A_{www}& =0, & A_{rww}& =0. & &
\end{align*}%
Integration of this linear system in partial derivatives is elementary,
because the most part of third derivatives vanishes. Straightforward
computation yields the list of local three-dimensional conservation laws
presented at the beginning of Section 2.

\section{Conclusion}

\label{sec:conc}

In this paper we established a new phenomenon in the theory of integrable
systems. We discovered a new property: an existence of infinitely many local
three-dimensional conservation laws for pairs of commuting two-dimensional
integrable systems. We considered two remarkable three-dimensional
integrable quasilinear systems of second order well known as the
dispersionless limit of the Kadomtsev--Petviashvili equation and the Mikhal%
\"{e}v equation. We show that their two-dimensional hydrodynamic reductions
(multi-component quasilinear systems of first order) have such local
three-dimensional hydrodynamic conservation laws. We constructed
two-parametric generating equations of three-dimensional hydrodynamic
conservation laws for these hydrodynamic reductions and for corresponding
hydrodynamic chains. The Mikhal\"{e}v equation also possesses infinitely
many two-dimensional dispersive reductions. Also in this paper we considered
the Korteweg de Vries pair of commuting flows as a simplest dispersive
reduction of the Mikhal\"{e}v equation. Infinitely many local
three-dimensional conservation laws were found for this pair of commuting
dispersive equations.

For simplicity we restricted our consideration to pairs of commuting flows,
whose Lax pairs are well known. Derivation of local three-dimensional
conservation laws and corresponding generating equations is based on the
theory of over-determined systems. Thus, one can consider any commuting pair
of two-dimensional integrable systems and find corresponding local
three-dimensional conservation laws.

Our motivation in this particular research was based on two important
reasons. In the theory of shock waves discontinuities in a three-dimensional
case can be interpreted as discontinuities of commuting two-dimensional
reductions, which possess infinitely many local two-dimensional and
three-dimensional conservation laws. Application of the Whitham averaging
approach to three-dimensional integrable dispersive systems leads to
three-dimensional quasilinear systems of first order, whose two-dimensional
hydrodynamic reductions possess infinitely many two-dimensional and
three-dimensional hydrodynamic conservation laws. We believe that our
observation is a first step in derivation of infinitely many
three-dimensional hydrodynamic conservation laws for such averaged systems.
This problem should be investigated in a separate publication.

Our first computations of three-dimensional hydrodynamic conservation laws
for the dispersionless limit of the Kadomtsev--Petviashvili equation show
that the concept of local conservation laws can be extended to quasilocal
conservation laws for three-dimensional integrable systems. At the first
stage of our investigation (see details in \cite{ZM}) we were able to prove
existence of infinitely many such quasilocal conservation laws for the
dispersionless limit of the Kadomtsev--Petviashvili equation, reducing
corresponding computations to infinitely many over-determined systems in
involutions, whose general solutions depend on arbitrary constants only.
Instead of infinitely many particular calculations, in this paper we solved
just one such an over-determined system, whose general solution determines a
generating function of three-dimensional quasilocal conservation laws. This
means that computation of infinitely many these conservation laws is reduced
to the expansions of generating equations with respect to two arbitrary
parameters.

The well known fact is that any integrable hydrodynamic chain possesses just
one infinite series of two-dimensional hydrodynamic conservation laws, while
its $N$ component hydrodynamic reductions have $N$ infinite series of
two-dimensional hydrodynamic conservation laws. In this paper we show that
pairs of commuting integrable hydrodynamic chains possess two-parametric
family of three-dimensional hydrodynamic conservation laws, while
corresponding $N$ component hydrodynamic reductions have $2N$ infinite
series of three-dimensional hydrodynamic conservation laws. In a separate
paper we are going to continue investigate other properties of
multi-dimensional hydrodynamic conservation laws for families of
two-dimensional commuting semi-Hamiltonian hydrodynamic type systems.

\section*{Acknowledgements}

ZVM would like to thank A.A.~Cherevko, E.V.~Ferapontov and N.I.~Makarenko
for helpful conversations. MVP is grateful also to V.E.~Adler, A.V.~Aks\"{e}%
nov, G.A.~Alekseev, L.V.~Bogdanov, I.S.~Krasil'shchik, A.G.~Kulikovskii,
A.Ya.~Maltsev, O.I.~Morozov, D.V.~Treschev, A.M.~Verbovetsky for very
important comments, remarks and advices. This work was partially supported
by the Russian Science Foundation (grant No. 15-11-20013).

\end{document}